\documentclass[final,5p,times,twocolumn,nopreprintline]{elsarticle}

\usepackage{graphicx}
\usepackage{dcolumn}
\usepackage{bm}
\usepackage{amsmath,amsfonts}
\usepackage{verbatim}
\usepackage[dvipsnames]{xcolor}
\usepackage{threeparttable}
\usepackage{booktabs}
\usepackage{epsfig}
\usepackage{stmaryrd}
\usepackage{amssymb,tabularx}
\usepackage{amstext}
\usepackage{float}
\usepackage{mathtools}
\usepackage{multirow}
\usepackage{makecell}
\usepackage{tabu}
\usepackage{braket}
\usepackage{subfigure}
\usepackage{enumitem}
\usepackage{placeins}
\usepackage[colorlinks=true,
bookmarks=true,
linkcolor=blue,
urlcolor=blue,
anchorcolor=black,
citecolor=blue]{hyperref}
\renewcommand{\vec}[1]{\mbox{\boldmath $#1$}}

\hypersetup{
pdftitle={A relativistic mechanism for the enhanced isovector spin-orbit interaction suggested by parity-violating electron scattering experiments},
pdfauthor={Mengying Qiu; Tong-Gang Yue; Zhen Zhang; Lie-Wen Chen}
}

\journal{Physics Letters B}
\biboptions{sort&compress}

\begin{document}

\begin{frontmatter}

\title{A relativistic mechanism for the enhanced isovector spin-orbit interaction suggested by parity-violating electron scattering experiments}

\author[inst1]{Mengying Qiu}

\author[inst2,inst3]{Tong-Gang Yue}

\author[inst1,inst4]{Zhen Zhang\corref{cor1}}
\ead{zhangzh275@mail.sysu.edu.cn}

\author[inst2]{Lie-Wen Chen\corref{cor2}}
\ead{lwchen@sjtu.edu.cn}

\cortext[cor1]{Corresponding author.}
\cortext[cor2]{Corresponding author.}

\address[inst1]{Sino-French Institute of Nuclear Engineering and Technology, Sun Yat-Sen University, Zhuhai 519082, China}
\address[inst2]{State Key Laboratory of Dark Matter Physics, Key Laboratory for Particle Astrophysics and Cosmology (MOE), and Shanghai Key Laboratory for Particle Physics and Cosmology, School of Physics and Astronomy, Shanghai Jiao Tong University, Shanghai 200240, China}
\address[inst3]{Tsung-Dao Lee Institute, Shanghai Jiao Tong University, Shanghai 201210, China}
\address[inst4]{Shanghai Research Center for Theoretical Nuclear Physics, NSFC and Fudan University, Shanghai 200438, China}

\begin{abstract}
Recent high-precision parity-violating electron scattering (PVES) measurements on $^{208}$Pb (PREX-II) and $^{48}$Ca (CREX) reveal a tension in their simultaneous description within modern nuclear energy density functionals (EDFs). Analyses of these data suggest that an enhanced isovector spin-orbit interaction may help account for both measurements, but its relativistic origin in covariant density functional theory remains to be clarified. We show that, within the framework of a covariant density-dependent point-coupling EDF, an enhanced isovector tensor coupling can naturally induce such a strong isovector spin-orbit interaction. This mechanism provides a promising route toward a simultaneous description of the PREX-II and CREX results while preserving a reasonable description of finite nuclei and nuclear matter. PVES on $^{48}$Ca thus provides a sensitive probe of the covariant isovector tensor interaction.
\end{abstract}


\end{frontmatter}

\section{Introduction}

Parity-violating electron scattering (PVES) provides a unique and largely model-independent probe of neutron distributions in nuclei~\cite{Donnelly:1989qs,Horowitz:1999fk}. By measuring the parity-violating asymmetry in polarized elastic electron scattering, one can extract the weak-charge density distribution and the closely related neutron density with minimal strong-interaction uncertainties. Such information on neutron and weak densities is highly relevant for a broad range of phenomena, including nuclear structure, isovector interactions, neutron-rich matter in astrophysics~\cite{Furnstahl:2001un,Yoshida:2004mv,Chen:2005ti,Centelles:2008vu,Chen:2010qx,Reinhard:2010wz,Roca-Maza:2011qcr,Agrawal:2012pq,Zhang:2013wna,Mondal:2016bls,Raduta:2017cma,Thiel:2019tkm,Newton:2020jwn,Reinhard:2021utv,Lynch:2021xkq}, nuclear weak-interaction processes~\cite{COHERENT:2017ipa,Cadeddu:2017etk,Huang:2023aob,Piekarewicz:2025lel}, and even dark matter detection~\cite{Zheng:2014nga}.

Two recent high-precision PVES experiments at Jefferson Lab, PREX-II~\cite{PREX:2021umo} and CREX~\cite{CREX:2022kgg}, have provided precise measurements of parity-violating asymmetries and the corresponding differences between the charge and weak form factors, $\Delta F_{\mathrm{CW}}(q)\equiv F_{\mathrm{C}}(q)-F_{\mathrm{W}}(q)$, in $^{208}$Pb and $^{48}$Ca at their respective momentum transfers. However, an apparent tension emerged in early attempts to describe both measurements simultaneously within modern nuclear energy density functional (EDF) theory~\cite{CREX:2022kgg}. In particular, conventional nuclear EDFs fail to reproduce the PREX-II and CREX data simultaneously within their $1\sigma$ uncertainties~\cite{CREX:2022kgg,Reinhard:2022inh,Zhang:2022bni,Yuksel:2022umn,Miyatsu:2023lki,Kumar:2023bmb}, suggesting that important ingredients in the isovector sector may be missing or insufficiently constrained.

Considerable efforts have been devoted to reconciling the PREX-CREX tension, but with limited success~\cite{Mondal:2022cva,Yang:2023pdr,Reed:2023cap,Salinas:2023qic,Roca-Maza:2025vnr}. Recently, a study based on extended Skyrme EDFs has shown that introducing a substantially enhanced isovector spin-orbit (IVSO) interaction can reconcile the PREX-II and CREX measurements by exploiting the distinct shell structures of $^{48}$Ca and $^{208}$Pb~\cite{Yue:2024srj} (see also~\cite{Zhao:2024gjz}). Both data sets can be reproduced within their $1\sigma$ uncertainties without compromising the overall description of basic nuclear global properties~\cite{Yue:2024srj}. In a complementary relativistic analysis, Ref.~\cite{Kunjipurayil:2025xss} recast the Dirac equation into an equivalent Schr\"odinger-like form to isolate the isovector component of the spin-orbit (SO) potential and demonstrated its strong impact on neutron distributions in $^{48}$Ca. These studies point to an enhanced IVSO interaction as a key ingredient and motivate the question of its covariant origin.

In covariant density functional theory, the spin-orbit coupling is not introduced as an independent term but naturally emerges from the Lorentz structure of the effective nuclear interaction via the nucleon Dirac equation. Tensor couplings, as additional Lorentz channels beyond the familiar scalar and vector ones, provide a further relativistic contribution to the SO splittings~\cite{Furnstahl:1997tk,Jiang:2005vu,Sulaksono:2011zz, Mao:2002xm}. Despite their fundamental role~\cite{Rufa:1988zz,Long:2007dw,Typel:2020ozc,Zhao:2022xhq,Salinas:2023qic,Mao:2002xm, Mercier:2022svw, Typel:2024myq}, the strengths of tensor couplings, especially the isovector tensor coupling (IVTC), remain poorly constrained. Only a few observables provide direct sensitivity to them and their effects are strongly correlated with scalar and vector mean fields. Consequently, tensor terms are commonly omitted or strongly restricted in most covariant EDFs.

In this Letter, we examine whether the PREX-II and CREX data can be understood in terms of an enhanced isovector tensor coupling in covariant density functional theory. Using density-dependent point-coupling (DDPC) functionals with explicit tensor terms, we find that an enhanced IVTC naturally generates a stronger IVSO interaction and offers a covariant route toward accommodating the PREX-II and CREX data simultaneously, while maintaining a reasonable description of finite nuclei and nuclear matter.

\section{Theoretical framework}

Our analysis is carried out within the framework of the covariant DDPC EDF, which describes the effective nucleon-nucleon interaction through local four-fermion contact terms in the isoscalar-scalar (S), isoscalar-vector (V), isovector-scalar ($\tau$S), and isovector-vector ($\tau$V) channels, supplemented by derivative terms that simulate leading finite-range effects~\cite{Niksic:2008vp}. In this work, we adopt the extended DDPC functional introduced in Ref.~\cite{Zhao:2022xhq}, where isoscalar and isovector tensor couplings are explicitly included through the contact terms
\begin{equation}
-\frac{1}{2}\alpha_{\mathrm{T}}(\bar{\psi}\sigma_{\mu\nu}\psi)^2
-\frac{1}{2}\alpha_{\tau\mathrm{T}}(\bar{\psi}\sigma_{\mu\nu}\vec{\tau}\psi)^2,
\end{equation}
with $\psi$ denoting the nucleon field, $\vec{\tau}$ the isospin Pauli matrices, and $\sigma_{\mu\nu}=\tfrac{i}{2}[\gamma_\mu,\gamma_\nu]$ the Dirac tensor operator. The coupling constants $\alpha_{\mathrm{T}}$ and $\alpha_{\tau\mathrm{T}}$ characterize the strengths of the isoscalar and isovector tensor couplings, respectively.

The remaining scalar, vector, isovector-scalar, and isovector-vector couplings $\alpha_i$ ($i=\mathrm{S},\,\tau\mathrm{S},\,\mathrm{V},\,\tau\mathrm{V}$) are taken to be density dependent and parametrized as $\alpha_i(\rho)=\alpha_i(\rho_{\mathrm{sat}})f_i(\rho/\rho_{\mathrm{sat}})$, with
\[
f_i(x)=a_i\frac{1+b_i(x+d_i)^2}{1+c_i(x+d_i)^2},
\]
and the saturation density $\rho_{\mathrm{sat}}=0.156~\mathrm{fm}^{-3}$. The constraints $f_i(1)=1$ and $f_i''(0)=0$ are imposed following Refs.~\cite{Typel:1999yq,Zhao:2022xhq}. 
The isoscalar tensor coupling $\alpha_{\mathrm{T}}$ and the derivative coupling $\delta_{\mathrm{S}}$ are taken to be density independent.  In Ref.~\cite{Zhao:2022xhq}, $\alpha_{\tau\mathrm{T}}$ is constrained
by Fierz relations, which map the exchange (Fock) contributions of
the underlying couplings onto effective direct terms and thereby
make $\alpha_{\tau\mathrm{T}}$ density dependent. In this work, we relax this constraint and treat $\alpha_{\tau\mathrm{T}}$ as a free, density-independent coupling within an extended Hartree-level DDPC framework. All finite-nucleus calculations are performed by solving the relativistic Hartree-Bogoliubov equations using a modified numerical implementation based on Ref.~\cite{Niksic:2014dra}, with pairing parameters taken from the PCF-PK1 EDF~\cite{Zhao:2022xhq}.

\begin{figure}[t]
\centering
\includegraphics[width=\linewidth]{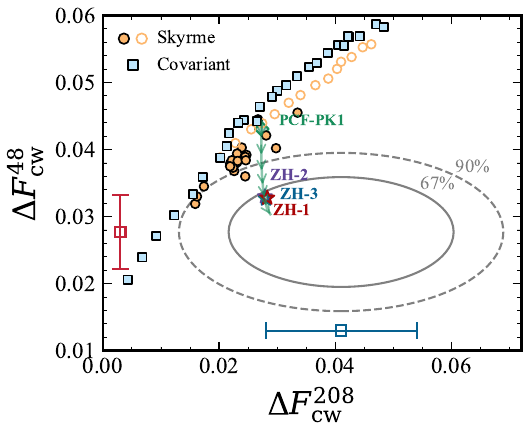}
\caption{Predicted charge-weak form-factor differences $\Delta F_{\mathrm{CW}}^{48}$ and $\Delta F_{\mathrm{CW}}^{208}$ for $^{48}\mathrm{Ca}$ and $^{208}\mathrm{Pb}$ from the ZH EDFs in comparison with the PREX-II and CREX results. A single diamond denotes the PCF-PK1 prediction~\cite{Zhao:2022xhq}, while squares and circles indicate other covariant EDFs~\cite{CREX:2022kgg} and nonrelativistic Skyrme-like EDFs~\cite{CREX:2022kgg,Yue:2021yfx}, respectively. The solid (dashed) ellipse denotes the joint 67\% (90\%) confidence region from the PREX-II and CREX measurements. Green arrows illustrate the evolution of the PCF-PK1 prediction as the IVTC $\alpha_{\tau\mathrm{T}}$ is varied from $0$ to $14~\mathrm{fm}^2$. Experimental values are $\Delta F_{\mathrm{CW}}^{208}(0.3977~\mathrm{fm}^{-1}) = 0.041 \pm 0.013$~\cite{PREX:2021umo} and $\Delta F_{\mathrm{CW}}^{48}(0.8733~\mathrm{fm}^{-1}) = 0.0277 \pm 0.0055$~\cite{CREX:2022kgg}.}
\label{fig:DFcw}
\end{figure}
\section{Results and discussion}
\subsection{PVES and the isovector tensor coupling}

We focus on the charge-weak form-factor differences $\Delta F_{\mathrm{CW}}(q)$ measured by PREX-II and CREX at their respective momentum transfers: $\Delta F_{\mathrm{CW}}^{208}(q=0.3977~\mathrm{fm}^{-1}) = 0.041\pm0.013$ for $^{208}$Pb~\cite{PREX:2021umo} and $\Delta F_{\mathrm{CW}}^{48}(q=0.8733~\mathrm{fm}^{-1}) = 0.0277\pm0.0055$ for $^{48}$Ca~\cite{CREX:2022kgg}. 
The calculation of the charge and weak-charge form factors entering $\Delta F_{\mathrm{CW}}$ is described in \ref{App:formfac}.
Fig.~\ref{fig:DFcw} compares the data with predictions from a representative set of covariant and nonrelativistic Skyrme-like EDFs~\cite{CREX:2022kgg,Yue:2021yfx} in the $\left(\Delta F_{\mathrm{CW}}^{48}, \Delta F_{\mathrm{CW}}^{208}\right)$ plane. As shown, predictions from these EDFs lie largely outside the 90\% joint confidence region reported by PREX-II and CREX, indicating that conventional EDFs cannot simultaneously describe both observables within their $1\sigma$ experimental uncertainties.

Starting from the well-calibrated DDPC functional PCF-PK1~\cite{Zhao:2022xhq}, we vary the IVTC strength $\alpha_{\tau\mathrm{T}}$ from 0 to $14~\mathrm{fm}^2$ while keeping all other parameters fixed. The resulting predictions (green arrows in Fig.~\ref{fig:DFcw}) trace a clear trend in the $\left(\Delta F_{\mathrm{CW}}^{48},\Delta F_{\mathrm{CW}}^{208}\right)$ plane, with $\Delta F_{\mathrm{CW}}^{48}$ decreasing significantly while $\Delta F_{\mathrm{CW}}^{208}$ changes only slightly. Consequently, the predictions move systematically toward the joint PREX-CREX confidence region. This behavior reveals the pronounced sensitivity of $\Delta F_{\mathrm{CW}}^{48}$ to the IVTC.

\begin{table}[t]
\centering
\begin{threeparttable}
\caption{Values of model parameters and predictions of the three new EDFs, ZH-1, ZH-2, and ZH-3, in comparison with PCF-PK1~\cite{Zhao:2022xhq}. Predictions are shown for the charge-weak form-factor difference $\Delta F_{\rm{CW}}$ and the neutron skin thickness $\Delta r_{\rm{np}}$ in $^{48}$Ca and $^{208}$Pb, as well as the symmetry energy $E_{\rm{sym}}(\rho)$ and its density slope parameter $L(\rho)$ at $\rho_{\rm{sat}}$ and $2\rho_{\rm{sat}}/3$. The parameters $\alpha_{\tau \mathrm{S}}(\rho_{\mathrm{sat}})$, $\alpha_{\tau \mathrm{V}}(\rho_{\mathrm{sat}})$, $\alpha_{\mathrm{T}}$, and $\alpha_{\tau \mathrm{T}}$ are in $\mathrm{fm}^2$, while $\delta_{\mathrm{S}}$ is in $\mathrm{fm}^4$. $\Delta r_{\mathrm{np}}$ values are in $\mathrm{fm}$; $E_{\mathrm{sym}}$ and $L$ are in $\mathrm{MeV}$. The superscripts ``48" and ``208" indicate results for $^{48}$Ca and $^{208}$Pb, respectively. All four EDFs share identical parameters in the isoscalar sector: $\alpha_{\mathrm{S}}(\rho_{\mathrm{sat}}) = -6.315~\mathrm{fm}^{2}$, $a_{\mathrm{S}} = 1.425 $, $b_{\mathrm{S}}=0.3185$, $c_{\mathrm{S}} = 0.5929$, $d_{\mathrm{S}}=0.7498$, $\alpha_{\mathrm{V}}(\rho_{\mathrm{sat}}) = 3.533~\mathrm{fm}^{2}$, $a_{\mathrm{V}} = 0.8076 $, $b_{\mathrm{V}}=0.06058$, $c_{\mathrm{V}} = 0.03693$, $d_{\mathrm{V}}=3.005$.}
\label{tab:param}
\small
\setlength{\tabcolsep}{5pt}
\begin{tabular}{lcccc}
\toprule
 & PCF-PK1 & ZH-1 & ZH-2 & ZH-3 \\
\midrule
$\alpha_{\tau \mathrm{S}}(\rho_{\mathrm{sat}})$ & -1.981 & -4.701 & -0.8403 & -3.321 \\
$a_{\tau \mathrm{S}}$ & 2.328 & 0.5947 & 2.126 & 1.802 \\
$b_{\tau \mathrm{S}}$ & 0.06179 & 84.914 & 334.1 & 9.290 \\
$c_{\tau \mathrm{S}}$ & 0.5704 & 50.16 & 711.5 & 17.36 \\
$d_{\tau \mathrm{S}}$ & 0.7644 & 0.08152 & 0.02165 & 0.1386 \\
$\alpha_{\tau \mathrm{V}}(\rho_{\mathrm{sat}})$ & 2.976 & 5.400 & 1.927 & 4.096 \\
$a_{\tau \mathrm{V}}$ & 2.547 & 4.212 & 9.430 & 6.656 \\
$b_{\tau \mathrm{V}}$ & 0.3459 & -0.001241 & -0.001728 & -0.002950 \\
$c_{\tau \mathrm{V}}$ & 1.612 & 0.0006678 & 0.0007385 & 0.001406 \\
$d_{\tau \mathrm{V}}$ & 0.4547 & 22.34 & 21.24 & 15.39 \\
$\delta_{\mathrm{S}}$ & -0.6658 & -0.7173 & -0.7886 & -0.7311 \\
$\alpha_{\mathrm{T}}$ & 3.374 & 4.312 & 5.673 & 4.546 \\
$\alpha_{\tau \mathrm{T}}$ & 0.535$^{a}$ & 6.967 & 9.195 & 7.114 \\
$\Delta F_{\mathrm{CW}}^{48}$ & {0.0431} & {0.0328} & {0.0330} & {0.0326} \\
$\Delta F_{\mathrm{CW}}^{208}$ & {0.0273} & {0.0283} & {0.0279} & {0.0281} \\
$\Delta r_{\mathrm{np}}^{48}$ & {0.175} & 0.116 & 0.128 & 0.127 \\
$\Delta r_{\mathrm{np}}^{208}$ & 0.183 & {0.187} & {0.189} & 0.191 \\
$E_{\mathrm{sym}}(\rho_{\mathrm{sat}})$ & 33.0 & 33.6 & 32.4 & 32.1 \\
$E_{\mathrm{sym}}(2\rho_{\mathrm{sat}}/3)$ & 24.5 & 25.4 & 25.8 & 26.9 \\
$L(\rho_{\mathrm{sat}})$ & 78.4 & 66.7 & 43.8 & 19.4 \\
$L(2\rho_{\mathrm{sat}}/3)$ & 50.4 & 54.2 & 50.2 & 49.0 \\
\bottomrule
\end{tabular}
\begin{tablenotes}
\footnotesize
\item[a] The maximum value of $\alpha_{\tau \mathrm{T}}$ in $^{48}$Ca obtained with PCF-PK1.
\end{tablenotes}
\end{threeparttable}
\end{table}

Motivated by this observation, we adopt a two-step procedure based on parameter-space sampling and subsequent selection for the DDPC functional extended by the tensor couplings. Since $\Delta F_{\mathrm{CW}}$ is largely insensitive to the isoscalar sector, $\alpha_{\mathrm{S}}(\rho)$ and $\alpha_{\mathrm{V}}(\rho)$ are fixed at their PCF-PK1 values. In the sampling step, the remaining isovector couplings, tensor couplings, and the derivative term are varied. In the selection step, the sampled parameter sets are screened by requiring consistency with the PREX-II and CREX data, selected ground-state observables, and empirical and \emph{ab initio} constraints on the nuclear matter equation of state. We select three representative parametrizations, denoted by ZH-1, ZH-2, and ZH-3, which exhibit distinct density dependences of the symmetry energy. Their parameters and corresponding predictions for $\Delta F_{\mathrm{CW}}^{48}$ and $\Delta F_{\mathrm{CW}}^{208}$ are summarized in Table~\ref{tab:param}. As shown in Fig.~\ref{fig:DFcw}, all three EDFs reproduce both $\Delta F_{\mathrm{CW}}^{48}$ and $\Delta F_{\mathrm{CW}}^{208}$ within their respective $1\sigma$ uncertainties. In all three EDFs, the IVTC $\alpha_{\tau\mathrm{T}}$ is enhanced relative to PCF-PK1, while bulk nuclear properties and the nuclear matter equation of state remain reasonably described. Details of the two-step parameter-space sampling and selection procedure and global performance are provided in \ref{App:GSexp} and \ref{App:FN_PNM}.

\subsection{Isovector spin-orbit interaction induced by tensor couplings}

To clarify the microscopic origin of the improved PREX-CREX description, we perform a nonrelativistic reduction of the covariant EDF (see \ref{App:nonrel} for details). The SO part of the reduced energy density can be written as
\begin{equation}
\mathcal{E}_{\mathrm{SO}} =
\frac{b_{\mathrm{IS}}}{2}{\nabla}\rho\cdot\bm{J}
+\frac{b_{\mathrm{IV}}}{2}{\nabla}\tilde{\rho}\cdot\bm{\tilde{J}}
+\frac{b_{\mathrm{SV}}}{2}{\nabla}\rho\cdot\bm{\tilde{J}},
\end{equation}
with
\begin{eqnarray}
    b_{\mathrm{IS}} &=& 8\mathcal{B}_0^2\,\alpha_{\mathrm{T}}
-4\mathcal{B}_0^2\,\left[\alpha_{\mathrm{S}}(\rho)+\alpha'_{\mathrm{S}}(\rho)\,\rho\right],\\
b_{\mathrm{IV}} &=& 8\mathcal{B}_0^2\,\alpha_{\tau\mathrm{T}}
-4\mathcal{B}_0^2\,\alpha_{\tau\mathrm{S}}(\rho),\\
b_{\mathrm{SV}} &=& -4\mathcal{B}_0^2\,\alpha'_{\tau\mathrm{S}}(\rho)\,\tilde{\rho}.
\end{eqnarray}
Here $\mathcal{B}_0 = 1/(2m^*)$, where $m^*$ denotes the isoscalar Dirac effective mass. In addition, $\rho=\rho_n+\rho_p$ ($\tilde{\rho}=\rho_n-\rho_p$) and $\bm{J}=\bm{J}_n+\bm{J}_p$ ($\bm{\tilde{J}}=\bm{J}_n-\bm{J}_p$) denote the isoscalar (isovector) densities and SO densities, respectively. Compared with the nonrelativistic Skyrme EDF, this expression shows explicit density dependences in the isoscalar and isovector SO coupling parameters $b_{\mathrm{IS}}$ and $b_{\mathrm{IV}}$, and includes an additional $b_{\mathrm{SV}}$ term arising from the density dependence of $\alpha_{\tau\mathrm{S}}$.

The density-weighted averages $\langle b_{\mathrm{IS}}\rangle$, 
$\langle b_{\mathrm{IV}}\rangle$, and $\langle b_{\mathrm{SV}}\rangle$ 
for $^{48}$Ca and $^{208}$Pb obtained with the ZH family and PCF-PK1 are 
summarized in Table~\ref{tab:so}. The mixed term $b_{\mathrm{SV}}$ is 
numerically small, with $|\langle b_{\mathrm{SV}}\rangle|$ not exceeding 
2.22~$\mathrm{MeV}\cdot\mathrm{fm}^5$ in PCF-PK1 and remaining below 
0.81~$\mathrm{MeV}\cdot\mathrm{fm}^5$ in the ZH EDFs. The dominant SO 
strengths are therefore $b_{\mathrm{IS}}$ and $b_{\mathrm{IV}}$. Relative 
to PCF-PK1, $\langle b_{\mathrm{IS}}\rangle$ in the ZH family increases 
only moderately, by a few tens of percent, whereas 
$\langle b_{\mathrm{IV}}\rangle$ is enhanced by a factor of about six.

\begin{figure}[t]
\centering
\includegraphics[width=\linewidth]{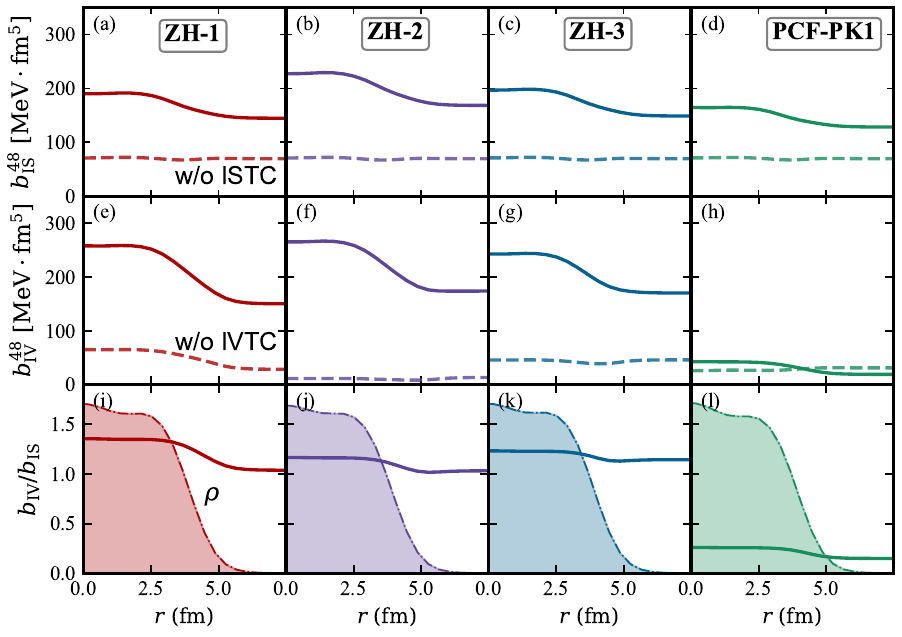}
\caption{Radial dependence of the isoscalar and isovector spin-orbit strengths, $b_{\mathrm{IS}}$ (top panels) and $b_{\mathrm{IV}}$ (middle panels), together with their ratio (bottom panels) in $^{48}$Ca, obtained from the nonrelativistic reduction of the ZH-1 (red), ZH-2 (purple), ZH-3 (blue), and PCF-PK1 (green) covariant EDFs. Dashed lines labeled by ``w/o ISTC'' and ``w/o IVTC'' denote the results obtained without the corresponding tensor terms. The shaded regions in the bottom panels indicate the nucleon density distributions in $^{48}$Ca.}
\label{fig:bisbiv}
\end{figure}

Figure~\ref{fig:bisbiv} displays the radial dependence of $b_{\mathrm{IS}}$, 
$b_{\mathrm{IV}}$, and their ratio $b_{\mathrm{IV}}/b_{\mathrm{IS}}$ in 
$^{48}$Ca for ZH-1, ZH-2, ZH-3, and PCF-PK1. The solid lines show the full 
results. To illustrate the contributions from tensor couplings, the dashed lines 
show $b_{\mathrm{IS}}$ calculated without the isoscalar tensor coupling 
(``w/o ISTC'') and $b_{\mathrm{IV}}$ calculated without the isovector tensor 
coupling (``w/o IVTC''). In the ZH family, the strong IVTC markedly enhances $b_{\mathrm{IV}}$, 
whereas the isoscalar tensor coupling only moderately increases $b_{\mathrm{IS}}$. Consequently, the ratio 
$b_{\mathrm{IV}}/b_{\mathrm{IS}}$ increases from about $1/3$ in PCF-PK1 to 
values above unity in the ZH EDFs over the nuclear interior and surface region. 
The results for $^{208}$Pb, not shown here, follow the same qualitative trend, 
with only minor quantitative differences.

\begin{table}[t]
\centering
\caption{Density-averaged isoscalar, isovector, and mixed spin-orbit coupling strengths $\langle b_{\mathrm{IS}}\rangle$, $\langle b_{\mathrm{IV}}\rangle$, and $\langle b_{\mathrm{SV}}\rangle$ (in units of $\mathrm{MeV}\cdot\mathrm{fm}^5$) in $^{48}$Ca and $^{208}$Pb, obtained from the nonrelativistic reduction of the covariant EDFs PCF-PK1~\cite{Zhao:2022xhq} and the three new parametrizations ZH-1, ZH-2, and ZH-3.}
\label{tab:so}
\small
\setlength{\tabcolsep}{3pt}
\renewcommand{\arraystretch}{0.95}
\begin{tabular}{lcccc}
\toprule
 & PCF-PK1 & ZH-1 & ZH-2 & ZH-3 \\
\midrule
$\langle b_{\text{IS}}^{48}\rangle$  & 150  & 172  & 205  & 178 \\
$\langle b_{\text{IS}}^{208}\rangle$ & 153  & 176  & 210  & 182 \\
$\langle b_{\text{IV}}^{48}\rangle$  & 35.0 & 222  & 230  & 213 \\
$\langle b_{\text{IV}}^{208}\rangle$ & 36.7 & 229  & 238  & 219 \\
$\langle b_{\text{SV}}^{48}\rangle$  & 1.48 & -0.302 & 0.0502 & 0.668 \\
$\langle b_{\text{SV}}^{208}\rangle$ & 2.22 & -0.349 & 0.0503 & 0.807 \\
\bottomrule
\end{tabular}
\end{table}

The improved description of the PREX-CREX data within the ZH family can therefore be traced to the strengthened IVSO interaction induced by the enhanced IVTC. Because the two nuclei have different shell structures, their responses to $b_{\mathrm{IV}}$ are markedly different. In spherical nuclei, orbitals with total angular momentum $j_{>}=l+\tfrac{1}{2}$ contribute positively to the corresponding nucleon SO density, whereas their partners with $j_{<}=l-\tfrac{1}{2}$ contribute negatively. In $^{208}\mathrm{Pb}$, the dominant spin-unsaturated contributions from the neutron $1i_{13/2}$ and proton $1h_{11/2}$ orbitals largely cancel in $\bm{\tilde{J}}=\bm{J}_n-\bm{J}_p$, leading to a weak sensitivity of $\Delta F_{\mathrm{CW}}^{208}$ to $b_{\mathrm{IV}}$. In contrast, in $^{48}\mathrm{Ca}$ all proton SO partners are filled, and the eight valence neutrons occupying the $1f_{7/2}$ orbital generate a large positive $J_n$, yielding a sizable $\bm{\tilde{J}}$ and an amplified response to $b_{\mathrm{IV}}$. Consistent with the nonrelativistic Skyrme-like EDF results~\cite{Yue:2024srj}, the enhanced $b_{\mathrm{IV}}$ in the ZH family markedly reduces $\Delta F_{\mathrm{CW}}^{48}$ while leaving $\Delta F_{\mathrm{CW}}^{208}$ nearly unchanged,
thereby bringing the predictions into consistency with both the PREX-II and CREX data within their $1\sigma$ uncertainties.

\subsection{Neutron skin thickness and the symmetry energy}

Having seen that the PREX-II and CREX $\Delta F_{\mathrm{CW}}$ data can be accommodated by an enhanced IVTC, we examine its impact on the neutron skin thicknesses $\Delta r_{\mathrm{np}}^{48}$ and $\Delta r_{\mathrm{np}}^{208}$ of $^{48}$Ca and $^{208}$Pb. As shown in Table~\ref{tab:param}, relative to PCF-PK1, the $\Delta r_{\mathrm{np}}^{48}$ predicted by the ZH EDFs decreases by about 30\%, whereas $\Delta r_{\mathrm{np}}^{208}$ changes only at the level of a few percent. This contrast is similar to the behavior of $\Delta F_{\mathrm{CW}}$. In $^{48}$Ca, the enhanced IVTC strengthens the IVSO coupling and modifies the neutron density distribution, reducing both $\Delta F_{\mathrm{CW}}^{48}$ and $\Delta r_{\mathrm{np}}^{48}$. In $^{208}$Pb, however, the small IVSO density $\tilde{J}$ results in a weak IVTC dependence of both observables.

Table~\ref{tab:param} also lists the symmetry energy $E_{\mathrm{sym}}(\rho)$ and its slope parameter $L(\rho)$ at $\rho_{\mathrm{sat}}$ and $2\rho_{\mathrm{sat}}/3$. Although $L(\rho_{\mathrm{sat}})$ varies widely across the ZH EDFs (19.4--66.7~MeV), $\Delta r_{\mathrm{np}}^{208}$ remains nearly unchanged, while $L(2\rho_{\mathrm{sat}}/3)$ varies little (49.0--54.2~MeV). This is consistent with the established correlation between $\Delta r_{\mathrm{np}}^{208}$ and $L(2\rho_{\mathrm{sat}}/3)$~\cite{Zhang:2013wna}, which remains robust even after incorporating a strong IVTC. In contrast, $\Delta r_{\mathrm{np}}^{48}$ depends not only on the density dependence of the symmetry energy but also directly on the IVTC strength, and therefore the ZH EDFs no longer follow the $\Delta r_{\mathrm{np}}^{48}$--$L(2\rho_{\mathrm{sat}}/3)$ trend obtained with conventional EDFs (see \ref{App:correlation}).

\section{Summary}
We have shown that an enhanced isovector tensor coupling within a covariant density-dependent point-coupling EDF offers a promising route toward a simultaneous description of the PREX-II and CREX PVES data. In the nonrelativistic limit, this coupling induces a stronger isovector spin-orbit interaction, providing a covariant interpretation of the enhanced isovector spin-orbit interaction suggested by the data while maintaining a reasonable description of nuclear global properties and the nuclear matter equation of state.

The enhanced isovector tensor coupling strongly suppresses $\Delta F_{\mathrm{CW}}^{48}$ in $^{48}$Ca while leaving $\Delta F_{\mathrm{CW}}^{208}$ in $^{208}$Pb nearly unchanged. This contrasting response reflects the distinct shell structures of the two nuclei and indicates that PVES on $^{48}$Ca is a sensitive electroweak probe of the isovector tensor sector in covariant energy density functionals.

More broadly, the present results suggest that precise PVES measurements constrain not only the density dependence of the symmetry energy but also the covariant isovector tensor interaction. The inferred strong IVTC is expected to influence the properties of neutron-rich nuclei, while the refined knowledge of neutron and weak-charge distributions in nuclei from PVES data has direct implications for neutron-rich matter and nuclear electroweak processes. The present two-step sampling-and-selection analysis should therefore be followed by a full statistical calibration over a broader EDF parameter space to quantify uncertainties and assess the robustness of the inferred IVTC under more extensive finite-nucleus and nuclear-matter constraints.

\FloatBarrier
\appendix
\setcounter{figure}{0}
\setcounter{table}{0}
\setcounter{equation}{0}
\numberwithin{figure}{section}
\numberwithin{table}{section}
\numberwithin{equation}{section}
\renewcommand{\thefigure}{\Alph{section}\arabic{figure}}
\renewcommand{\thetable}{\Alph{section}\arabic{table}}
\renewcommand{\theequation}{\Alph{section}\arabic{equation}}
\renewcommand{\theHfigure}{appendix.\thefigure}
\renewcommand{\theHtable}{appendix.\thetable}
\renewcommand{\theHequation}{appendix.\theequation}
\FloatBarrier

\section{Calculations of Charge and Weak-Charge Form Factors\label{App:formfac}}
In the framework of covariant energy density functionals (EDFs), the charge and weak-charge form factors are calculated following the prescription of Ref.~\cite{Horowitz:2012we}, where contributions from both the vector densities and tensor currents are explicitly included. Within the impulse approximation, the charge form factors $F_{\mathrm{C}}$ and weak-charge form factors $F_{\mathrm{W}}$, both normalized to one at $q=0$, are given by:
\begin{subequations}\label{eq:formfac}
\begin{align}
F_{\mathrm{C}}(q)=&
\frac{e^{a_{\mathrm{cm}} q^2}}{Z}
\sum_{i=p, n}\Biggl\{
G_{\mathrm{E}}^i\left(q^2\right) F_{\mathrm{V}}^i(q)
\notag\\
&+\left(\frac{G_{\mathrm{M}}^i\left(q^2\right)-G_{\mathrm{E}}^i\left(q^2\right)}
{1+\lambda}\right)
\times\left[\lambda F_{\mathrm{V}}^i(q)+\frac{q}{2 m} F_{\mathrm{T}}^i(q)\right]
\Biggr\}, \label{eq:formfac_charge}\\
F_{\mathrm{W}}(q)=&
\frac{e^{a_{\mathrm{cm}} q^2}}{Z g_{\mathrm{v}}^{\mathrm{p}}+N g_{\mathrm{v}}^{\mathrm{n}}}
\sum_{i=p, n}\Biggl\{
\widetilde{G}_{\mathrm{E}}^i\left(q^2\right) F_{\mathrm{V}}^i(q)
\notag\\
&+\left(\frac{\widetilde{G}_{\mathrm{M}}^i\left(q^2\right)-\widetilde{G}_{\mathrm{E}}^i\left(q^2\right)}
{1+\lambda}\right)
\times\left[\lambda F_{\mathrm{V}}^i(q)+\frac{q}{2 m} F_{\mathrm{T}}^i(q)\right]
\Biggr\}. \label{eq:formfac_weak}
\end{align}
\end{subequations}
where $\lambda=q^2/(4m^2)$, and $a_{\mathrm{cm}} = \hbar^2 \big/ \left( 8 \langle \hat{P}_{\mathrm{c.m.}}^2 \rangle \right)
$ accounts for the center-of-mass correction. $G_{\mathrm{E,M}}^i$ are the intrinsic nucleon electromagnetic form factors, and $\widetilde{G}_{\mathrm{E,M}}^i$ are the weak intrinsic nucleon form factors. $g_{\mathrm{v}}^{\mathrm{n}}$ and $g_{\mathrm{v}}^{\mathrm{p}}$ are the radiatively corrected weak-vector charges of the neutron and proton, respectively.

Under spherical symmetry, the single-particle solution of the Dirac equation takes the form
\begin{equation}
\begin{aligned}
\psi_{n \kappa m}(\mathbf{r})=\frac{1}{r}\binom{g_{n \kappa}(r) \mathcal{Y}_{+\kappa m}(\hat{\mathbf{r}})}{i f_{n \kappa}(r) \mathcal{Y}_{-\kappa m}(\hat{\mathbf{r}})}
\end{aligned}
\end{equation}
where $n$, $\kappa$, and $m$ are the principal, Dirac, and magnetic quantum numbers, respectively, and $\mathcal{Y}_{\pm \kappa m}(\hat{\mathbf{r}})$ are the spin spherical harmonics. The vector and tensor form factors are then expressed as
\begin{equation}
\begin{aligned}
 F_{\mathrm{V}}(q)=&\int \bar{\psi}(\mathbf{r}) e^{i \mathbf{q} \cdot \mathbf{r}} \gamma^0 \psi(\mathbf{r}) d^3 r\\
=& \sum_{n \kappa}(2 j+1) \int_0^{\infty}\left[g_{n \kappa}^2(r)+f_{n \kappa}^2(r)\right] j_0(q r) d r,\\
 F_{\mathrm{T}}(q)=&\int \bar{\psi}(\mathbf{r}) e^{i \mathbf{q} \cdot \mathbf{r}} \gamma^0 \boldsymbol{\gamma} \cdot \hat{\mathbf{q}} \psi(\mathbf{r}) d^3 r \\
 =&\sum_{n \kappa} 2(2 j+1) \int_0^{\infty} g_{n \kappa}(r) f_{n \kappa}(r) j_1(q r) d r,
\end{aligned}
\end{equation}
where $j_{l}(qr)$ are spherical Bessel functions. 

For intrinsic single-nucleon electromagnetic form factors, a simple dipole parametrization is adopted, which provides a good approximation at moderate momentum transfers. Specifically,
\begin{equation}
G_{\mathrm{E}}^p\left(q^2\right)=\frac{G_{\mathrm{M}}^p\left(q^2\right)}{\mu_p}=\frac{G_{\mathrm{M}}^n\left(q^2\right)}{\mu_n}=G_{\mathrm{D}}\left(q^2\right),
\end{equation}
where the dipole form factor is given by
\begin{equation}
G_{\mathrm{D}}\left(q^2\right)=\left(1+\frac{q^2}{12} r_p^2\right)^{-2}.
\end{equation}
The proton mean-square radius $r_p^2$, as well as the magnetic moments $\mu_p$ and $\mu_n$, are listed in Table~\ref{tab:formfac}. For the neutron electromagnetic form factor, we adopt the Galster parametrization
\begin{equation}
G_{\mathrm{E}}^n\left(q^2\right)=-\left(\frac{q^2 r_n^2 / 6}{1+q^2 / m^2}\right) G_{\mathrm{D}}\left(q^2\right),
\end{equation}
 where $r_{n}^2$ is the neutron mean-square radius.

As for weak intrinsic nucleon form factors, they can be expressed as the sum of the corresponding electromagnetic form factors, weighted by the nucleonic weak charges $g_{\rm v}^i$, together with a strange quark contribution, i.e.,
\begin{equation}
\begin{aligned}
&\widetilde{G}_{\mathrm{E}, \mathrm{M}}^p\left(q^2\right)=g_{\mathrm{v}}^{\mathrm{p}} G_{\mathrm{E}, \mathrm{M}}^p\left(q^2\right)+g_{\mathrm{v}}^{\mathrm{n}} G_{\mathrm{E}, \mathrm{M}}^n\left(q^2\right)+\xi_{\mathrm{v}}^{(0)} G_{\mathrm{E}, \mathrm{M}}^s\left(q^2\right),\\
&\widetilde{G}_{\mathrm{E}, \mathrm{M}}^n\left(q^2\right)=g_{\mathrm{v}}^{\mathrm{n}} G_{\mathrm{E}, \mathrm{M}}^p\left(q^2\right)+g_{\mathrm{v}}^{\mathrm{p}} G_{\mathrm{E}, \mathrm{M}}^n\left(q^2\right)+\xi_{\mathrm{v}}^{(0)} G_{\mathrm{E}, \mathrm{M}}^s\left(q^2\right),
\end{aligned}
\end{equation}
where $\xi_{\mathrm{v}}^{(0)}$ is the \textit{singlet} weak charge. For the strange quark electric form factor $G_{\mathrm{E}}^{s}$, we follow the parametrization of Ref.~\cite{Reinhard:2021utv} and adopt
\begin{equation}
\begin{aligned}
G_{\mathrm{E}}^s\left(q^2\right)&=\rho_s \frac{q^2 /\left(4 m^2\right)}{1+4.97 q^2 /\left(4 m^2\right)},
\end{aligned}
\end{equation}
where $\rho_s$ is the strange quark electric coupling. For the strange quark magnetic form factor, which should equal the strange magnetic moment $\kappa_s$ at $q^2=0$~\cite{Horowitz:2012we}, we adopt the following dipole-like parametrization
\begin{equation}
\begin{aligned}
G_{\mathrm{M}}^s\left(q^2\right)&=\frac{\kappa_{s}}{\left(1+\frac{q^2}{12 \kappa_{\mathrm{s}}} r_{\mathrm{M}, \mathrm{s}}^2\right)^2},
\end{aligned}
\end{equation}
where $r_{\mathrm{M}, \mathrm{s}}^2$ is the strange magnetic radius. All parameters entering the calculations of the charge and weak-charge form factors are listed in Table~\ref{tab:formfac}.

\begin{table*}[htbp]
\renewcommand{\arraystretch}{1.2} 
\setlength{\tabcolsep}{10pt}
\centering
\caption[Parameters used in charge and weak-charge form-factor calculations]{Parameters used in the calculation of charge and weak-charge form factors, including the radiatively corrected neutron and proton weak charges ($g_{\mathrm{v}}^{\mathrm{n}}$, $g_{\mathrm{v}}^{\mathrm{p}}$), nucleon mean-square radii ($r_p^2$, $r_n^2$), magnetic dipole moments ($\mu_p$, $\mu_n$), the singlet weak charge $\xi_{\mathrm{v}}^{(0)}$, the strange electric coupling $\rho_s$, the strange magnetic moment $\kappa_s$, and the strange magnetic radius $r_{\mathrm{M},s}^2$.
}\label{tab:formfac}
\begin{tabular}{ccc}
\hline \hline
parameters & values & Ref\\
\hline
 $g_{\mathrm{v}}^{\mathrm{p}}$ & 0.0721  &  \cite{Horowitz:2012we}\\
 $g_{\mathrm{v}}^{\mathrm{n}}$ & -0.9878 &\cite{Horowitz:2012we}\\
 $ r_p^2~[\mathrm{fm}^2]$ &  0.6906 & \cite{Xiong:2019umf} \\
$ r_n^2~[\mathrm{fm}^2]$ & -0.116& \cite{ParticleDataGroup:2018ovx} \\
 $\mu_p$ &  2.793 & \cite{Horowitz:2012we} \\
$\mu_n$ & -1.913 & \cite{Horowitz:2012we} \\
 $\xi_{\mathrm{v}}^{(0)}$& -0.9877  & \cite{Horowitz:2012we}\\
$\rho_s$ & -0.24  & \cite{Reinhard:2021utv} \\
 $\kappa_s$ & -0.017 &  \cite{Reinhard:2021utv}\\
$r_{\mathrm{M,s}}^2~[\mathrm{fm}^2]$& -0.015 &\cite{Alexandrou:2019olr}\\
\hline\hline
\end{tabular}
\end{table*}

\section{Sampling and selection protocol\label{App:GSexp}}
 In addition to the charge-weak form factor differences in $^{48}$Ca and $^{208}$Pb, the two-step sampling and selection procedure uses
 a selected set of experimental and empirical data to identify parametrizations with a reasonable description of both finite nuclei and nuclear matter.
Specifically, the nuclear ground-state data include the binding energies $E_{\mathrm{B}}$, charge radii $R_{\mathrm{C}}$ of $^{48}$Ca and $^{208}$Pb, and spin-orbit splittings $\Delta \epsilon_{\mathrm{ls}}$ of the $\nu 1f$ orbital in $^{48}$Ca and the $\pi 2 d$, $\nu 3p$, and $\nu 2f$ orbitals in $^{208}$Pb~\cite{Klupfel:2008af, Schwierz:2007ve, Isakov:2002jv}.
In the two-step sampling and selection protocol,
 we also employ the symmetry energy value at $2/3$ saturation 
density, 
$E_{\mathrm{sym}}(2\rho_{\mathrm{sat}}/3)=25.6^{+1.4}_{-1.3}~\mathrm{MeV}$, obtained from EDF analyses of nuclear masses~\cite{Qiu:2023kfu},
together with the neutron matter equation of state around $1/3$ saturation density,
$E_{\mathrm{PNM}}(\rho = 0.04805~\mathrm{fm}^{-3})=6.72\pm0.08~\mathrm{MeV}$ predicted by chiral effective field theory~\cite{Machleidt:2024bwl}.
The data used to guide the sampling and selection are summarized in Table~\ref{tab:GSexp}.

\begin{table*}[htbp]
\renewcommand{\arraystretch}{1.6} 
\setlength{\tabcolsep}{10pt}
\centering
\caption[Ground-state and nuclear-matter constraints used in the sampling and selection]{Experimental data for the ground-state properties of $^{48}$Ca and $^{208}$Pb, namely binding energies $E_{\mathrm{B}}$, charge radii $R_{\mathrm{C}}$, SO splittings $\Delta \epsilon_{\mathrm{ls}}$, and charge-weak form-factor differences $\Delta F_{\mathrm{CW}}$, together with constraints on the nuclear equation of state, used to guide the parameter-space sampling and selection. }\label{tab:GSexp}
\begin{tabular}{ccccc}
\hline \hline
\multicolumn{5}{c}{\textit{Finite nuclei constraints}} \\
\hline
Nucleus & $E_{\mathrm{B}}$ [MeV]~\cite{Klupfel:2008af} & $R_{\mathrm{C}}$ [fm]~\cite{Klupfel:2008af}& $\Delta \epsilon_{\mathrm{ls}}$ [MeV]~\cite{Schwierz:2007ve} & $\Delta F_{\mathrm{CW}}$~\cite{CREX:2022kgg, PREX:2021umo} \\
\hline
$^{48}$Ca & -415.990 & 3.477 & 8.80 ($\nu 1f$) &  0.0277 $\pm$ 0.0055\\ \hline
\multirow{3}{*}{$^{208}$Pb}  
          & \multirow{3}{*}{-1636.446} 
          & \multirow{3}{*}{5.501} 
          & 1.46 $(\pi 2d)$ \\
          &              &       & 0.89 $(\nu 3p)$ &  0.041 $\pm$ 0.013\\
          &              &       & 2.13 $(\nu 2f)$ \\
\hline 
\multicolumn{5}{c}{\textit{Nuclear matter constraints}} \\
\hline
& \multicolumn{2}{c}{$E_{\mathrm{sym}}(2\rho_{\mathrm{sat}}/3)\,[\mathrm{MeV}]$} &  \multicolumn{2}{c}{$E_{\mathrm{PNM}}(\rho = 0.04805~\mathrm{fm}^{-3})\,[\mathrm{MeV}]$}\\
\hline
&\multicolumn{2}{c}{$25.6^{+1.4}_{-1.3}$~\cite{Qiu:2023kfu}} &  \multicolumn{2}{c}{$6.72 \pm 0.08$~\cite{Machleidt:2024bwl}} \\
\hline\hline
\end{tabular}
\end{table*}

\begin{figure*}[tbp]
	\centering 
	\includegraphics[width=0.7\textwidth]{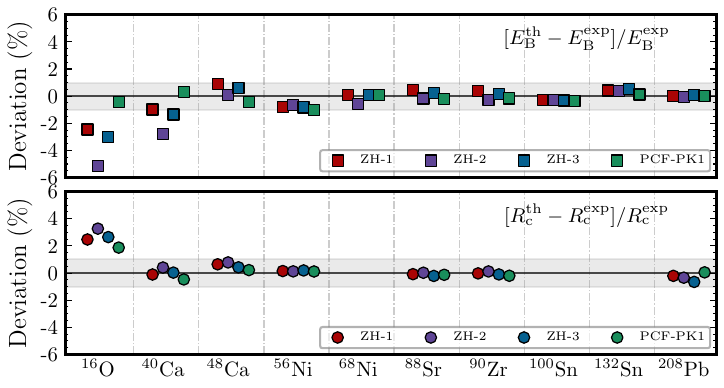}	
	\caption[Relative deviations of binding energies and charge radii]{Relative deviations of binding energies $E_{\mathrm{B}}$~(open squares)
and charge radii $R_{\mathrm{C}}$~(open circles) of $^{16}$O, $^{40}$Ca, $^{48}$Ca, $^{56}$Ni, $^{68}$Ni, $^{88}$Sr, $^{90}$Zr, $^{100}$Sn, $^{132}$Sn, and $^{208}$Pb predicted by ZH-1, ZH-2, and ZH-3 with respect to experimental measurements~\cite{Klupfel:2008af, Wang:2021xhn, Angeli:2013epw, LeBlanc:2005ik}. The shaded band represents a relative deviation of $\pm 1\%$.}
	\label{fig:BE_RC}%
\end{figure*}

\begin{figure*}[tbp]
	\centering 
	\includegraphics[width=0.6\textwidth]{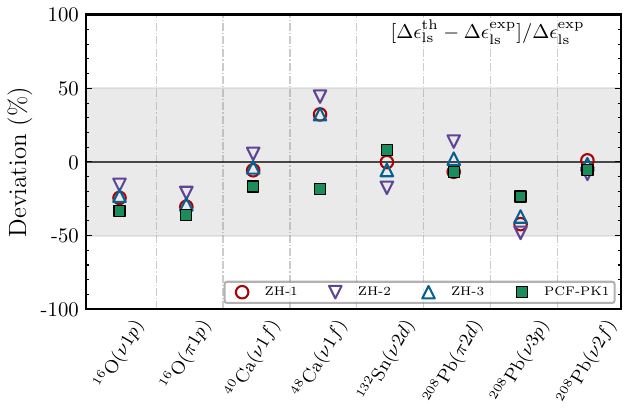}	
	\caption[Relative deviations of SO splittings]{Relative deviations of SO splittings $\Delta \epsilon_{\mathrm{ls}}$ in $^{16}$O, $^{40,48}$Ca, $^{132}$Sn, and $^{208}$Pb predicted by ZH-1, ZH-2, ZH-3, and PCF-PK1 with respect to experimental measurements~\cite{Schwierz:2007ve, Isakov:2002jv}. The shaded band represents a relative deviation of $\pm 50\%$.}
	\label{fig:Eso}%
\end{figure*}

Based on the selected dataset, we perform sampling using the emcee sampler~\cite{Foreman-Mackey:2012any} in parameter space according to the log-likelihood function
\begin{equation}\label{eq:loglike}
\log \mathcal{L}=-\sum_{i} \log \sigma_i-\frac{1}{2} \sum_{i} \frac{\left[\mathcal{O}_i^{\mathrm{exp}}-\mathcal{O}_i(\boldsymbol{\theta})\right]^2}{\sigma_i^2},
\end{equation}
where $\mathcal{O}_i^{\mathrm{exp}}$ denotes the experimental data, $\mathcal{O}_i(\boldsymbol{\theta})$ the theoretical predictions at parameter set $\boldsymbol{\theta}$, and $\sigma_i$ the adopted error for each type of observable. 
The $\sigma_i$ for binding energies, charge radii, and the relative errors of SO splittings are treated as nuisance hyperparameters associated with these data classes and sampled together with the EDF parameters, whereas those of the charge-weak form factor differences are fixed to their experimental values.

\begin{table*}[tbp]
\centering
\caption[Values of the nuisance hyperparameters for the selected ZH parametrizations]{Values of the nuisance hyperparameters entering Eq.~\ref{eq:loglike} for binding energies $E_{\mathrm{B}}$, charge radii $R_{\mathrm{C}}$, and the relative errors of SO splittings $\Delta \epsilon_{\mathrm{ls}}$ for the selected ZH parametrizations.}
\label{tab:sigma-parameters}
\begin{tabular}{lccc}
\hline\hline
Parametrization & $\sigma_{E_{\mathrm{B}}}~[\rm{MeV}]$ & $\sigma_{R_{\mathrm{C}}}~[\rm{fm}]$ & $\sigma_{\Delta \epsilon_{\mathrm{ls}}}/\Delta \epsilon_{\mathrm{ls}}^{\mathrm{exp}}$\\
\hline
ZH-1 & 3.57  & 0.233 & $24.8\%$ \\
ZH-2 & 1.46 & 0.038 & $30.5\%$ \\
ZH-3 & 2.14 & 0.067 & $26.0\%$ \\
\hline\hline
\end{tabular}
\end{table*}

From the full ensemble of samples, three new parametrizations are selected according to the following criteria:
(1) they simultaneously reproduce the charge-weak form factor differences measured by CREX and PREX-II within their $1\sigma$ uncertainties;
(2) the density slope of the symmetry energy at saturation density, $L(\rho_{\mathrm{sat}})$, lies in the range 10--100~MeV;
(3) the binding energies of $^{16}$O, $^{40}$Ca, $^{48}$Ca, $^{56}$Ni, $^{68}$Ni, $^{88}$Sr, $^{90}$Zr, $^{100}$Sn, $^{132}$Sn, and $^{208}$Pb, together with the charge radii of $^{16}$O, $^{40}$Ca, $^{48}$Ca, $^{56}$Ni, $^{88}$Sr, $^{90}$Zr, and $^{208}$Pb, are reproduced within approximately $1\%$; 
Finally, three parametrizations exhibiting distinct values of $L(\rho_{\mathrm{sat}})$ are chosen as the representative ZH family. The EDF parameters of these ZH parametrizations are summarized in the main text, while the corresponding values of the nuisance hyperparameters $\sigma_i$ are listed in Table~\ref{tab:sigma-parameters}.

\section{Performance of the new EDFs\label{App:FN_PNM}}
We begin by examining the performance of the three new EDFs for global properties of finite nuclei. The pairing channel is treated using the same separable finite-range Gogny D1S interaction and pairing parameters as in PCF-PK1~\cite{Zhao:2022xhq}. The ground-state properties of selected nuclei are calculated by  solving the relativistic Hartree-Bogoliubov equations in a harmonic-oscillator basis~\cite{Niksic:2014dra}, using $N_f=30$ major oscillator shells~\cite{Zhao:2022xhq}.
Fig.~\ref{fig:BE_RC} shows the relative deviations of the binding energies $E_{\mathrm{B}}$ and charge radii $R_{\mathrm{C}}$ from experimental data~\cite{Klupfel:2008af, Wang:2021xhn, Angeli:2013epw, LeBlanc:2005ik} for 10 typical doubly-magic and semi-magic nuclei. To account for the non-negligible finite nucleon size effects and intrinsic electromagnetic structure corrections~\cite{Xie:2025jmr}, the charge radii are computed from the corresponding charge form factor at zero momentum transfer as
\begin{equation}
R_{\mathrm{C}}=\sqrt{-\left.\frac{6}{F_{\mathrm{C}}(0)}\frac{d F_{\mathrm{C}}(q)}{d q^2}\right|_{q=0}}. 
\end{equation}
Although only properties of $^{48}$Ca and $^{208}$Pb were considered in the sampling and selection protocol, the new EDFs provide reasonable descriptions across the ten nuclei. The relative deviations generally lie within $1\%$, except for the light nuclei $^{16}$O and $^{40}$Ca.
Fig.~\ref{fig:Eso} shows the relative deviations of the SO splittings $\Delta \epsilon_{\mathrm{ls}}$ in $^{16}$O, $^{40,48}$Ca, $^{132}$Sn, and $^{208}$Pb with respect to experimental values~\cite{Schwierz:2007ve, Isakov:2002jv}. The ZH EDFs reproduce the experimental SO splittings with relative deviations typically below 50\%, without additional readjustment of the isoscalar sector. Detailed values for these observables predicted by the ZH family are listed in Table~\ref{tab:finite-nucleus-quality} in comparisons with the experimental values.

\begin{table*}[tbp]
\centering
\caption[Benchmark finite-nucleus and nuclear-matter observables]{Benchmark finite-nucleus and nuclear-matter observables predicted by the selected ZH-1, ZH-2, and ZH-3 parametrizations, compared with empirical or experimental reference values. Asterisks mark the data used to guide the two-step sampling and selection procedure.}
\label{tab:finite-nucleus-quality}
\footnotesize
\setlength{\tabcolsep}{5pt}
\renewcommand{\arraystretch}{1.05}
\begin{tabular}{@{}llcccc@{}}
\hline \hline
Observable & Nucleus/quantity & ZH-1 & ZH-2 & ZH-3 & Empirical/Exp. \\
\hline

\multirow{10}{*}{$E_{\mathrm{B}}$ [MeV]}
 & $^{16}$O   & -124.490 & -121.045 & -123.768 & -127.620 \\
 & $^{40}$Ca  & -338.681 & -332.513 & -337.408 & -342.051 \\
 & $^{48}$Ca  & -419.809 & -416.309 & -418.558 & -415.990\textsuperscript{*} \\
 & $^{56}$Ni  & -479.977 & -480.814 & -479.896 & -483.900 \\
 & $^{68}$Ni  & -590.981 & -586.974 & -590.974 & -590.430 \\
 & $^{88}$Sr  & -771.956 & -767.143 & -770.357 & -768.467 \\
 & $^{90}$Zr  & -786.830 & -781.736 & -785.082 & -783.893 \\
 & $^{100}$Sn & -823.512 & -823.325 & -823.190 & -825.800 \\
 & $^{132}$Sn & -1107.561 & -1107.080 & -1108.430 & -1102.900 \\
 & $^{208}$Pb & -1636.749 & -1635.360 & -1637.962 & -1636.446\textsuperscript{*} \\

\hline
\multirow{10}{*}{$R_{\mathrm{c}}$ [fm]}
 & $^{16}$O   & 2.768 & 2.789 & 2.773 & 2.701 \\
 & $^{40}$Ca  & 3.474 & 3.492 & 3.479 & 3.478 \\
 & $^{48}$Ca  & 3.499 & 3.504 & 3.492 & 3.477\textsuperscript{*} \\
 & $^{56}$Ni  & 3.755 & 3.755 & 3.757 & 3.750 \\
 & $^{68}$Ni  & 3.855 & 3.863 & 3.849 & -- \\
 & $^{88}$Sr  & 4.216 & 4.221 & 4.211 & 4.220 \\
 & $^{90}$Zr  & 4.268 & 4.274 & 4.265 & 4.269 \\
 & $^{100}$Sn & 4.479 & 4.480 & 4.481 & -- \\
 & $^{132}$Sn & 4.696 & 4.690 & 4.674 & -- \\
 & $^{208}$Pb & 5.490 & 5.482 & 5.465 & 5.501\textsuperscript{*} \\

\hline
\multirow{2}{*}{$\Delta F_{\mathrm{CW}}$}
 & $^{48}$Ca  & 0.0328 & 0.0330 & 0.0326 & $0.0277\pm0.0055$\textsuperscript{*} \\
 & $^{208}$Pb & 0.0283 & 0.0279 & 0.0281 & $0.041\pm0.013$\textsuperscript{*} \\

\hline
\multirow{8}{*}{$\Delta \epsilon_{\mathrm{ls}}$ [MeV]}
 & $^{16}$O $(\nu 1p)$     & 4.662  & 5.208  & 4.755  & 6.170 \\
 & $^{16}$O $(\pi 1p)$     & 4.395  & 4.987  & 4.500  & 6.320 \\
 & $^{40}$Ca $(\nu 1f)$    & 6.415  & 7.175  & 6.551  & 6.800 \\
 & $^{48}$Ca $(\nu 1f)$    & 11.640 & 12.703 & 11.653 & 8.800\textsuperscript{*} \\
 & $^{132}$Sn $(\nu 2d)$   & 1.658  & 1.368  & 1.569  & 1.660 \\
 & $^{208}$Pb $(\pi 2d)$   & 1.362  & 1.661  & 1.492  & 1.460\textsuperscript{*} \\
 & $^{208}$Pb $(\nu 3p)$   & 0.514  & 0.461  & 0.560  & 0.890\textsuperscript{*} \\
 & $^{208}$Pb $(\nu 2f)$   & 2.152  & 1.956  & 2.097  & 2.130\textsuperscript{*} \\

\hline
\multirow{2}{*}{EOS [MeV]}
 & $E_{\mathrm{sym}}(2\rho_{\mathrm{sat}}/3)$ 
 & 25.4 & 25.8 & 26.9 & $25.6^{+1.4}_{-1.3}$\textsuperscript{*} \\

 & $E_{\mathrm{PNM}}(\rho=0.04805~\mathrm{fm}^{-3})$ 
 & 6.66 & 6.78 & 6.74 & $6.72\pm0.08$\textsuperscript{*} \\

\hline \hline
\end{tabular}
\end{table*}

\begin{figure*}[tbp]
	\centering 
	\includegraphics[width=1.0\textwidth]{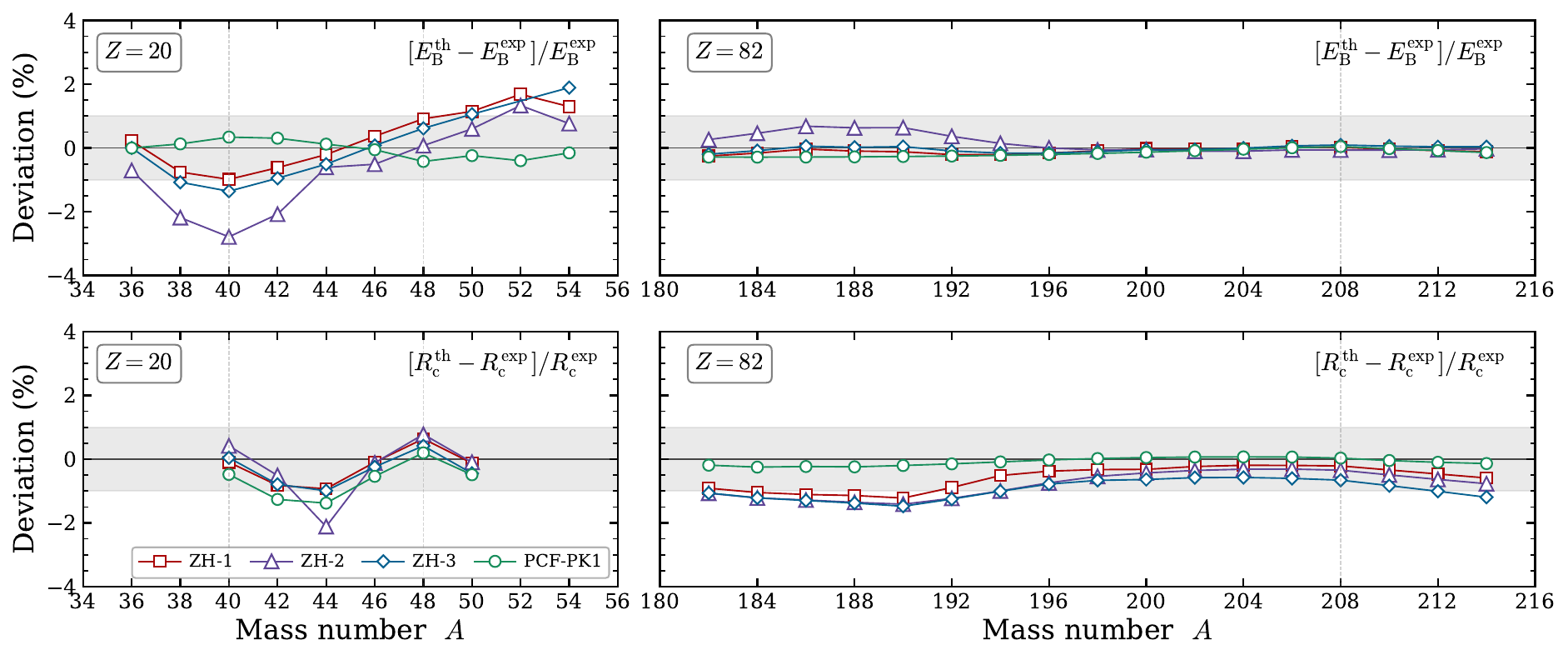}	
	\caption[Relative deviations along Ca and Pb isotopic chains]{Relative deviations of binding energies $E_{\mathrm{B}}$
and charge radii $R_{\mathrm{C}}$ of Ca and Pb isotopic chains with respect to experimental measurements~\cite{Klupfel:2008af, Wang:2021xhn, Angeli:2013epw, LeBlanc:2005ik}. The shaded band represents a relative deviation of $\pm 1\%$.}
	\label{fig:isotopes}%
\end{figure*}

\begin{figure*}[tbp]
	\centering 
	\includegraphics[width=1.0\textwidth]{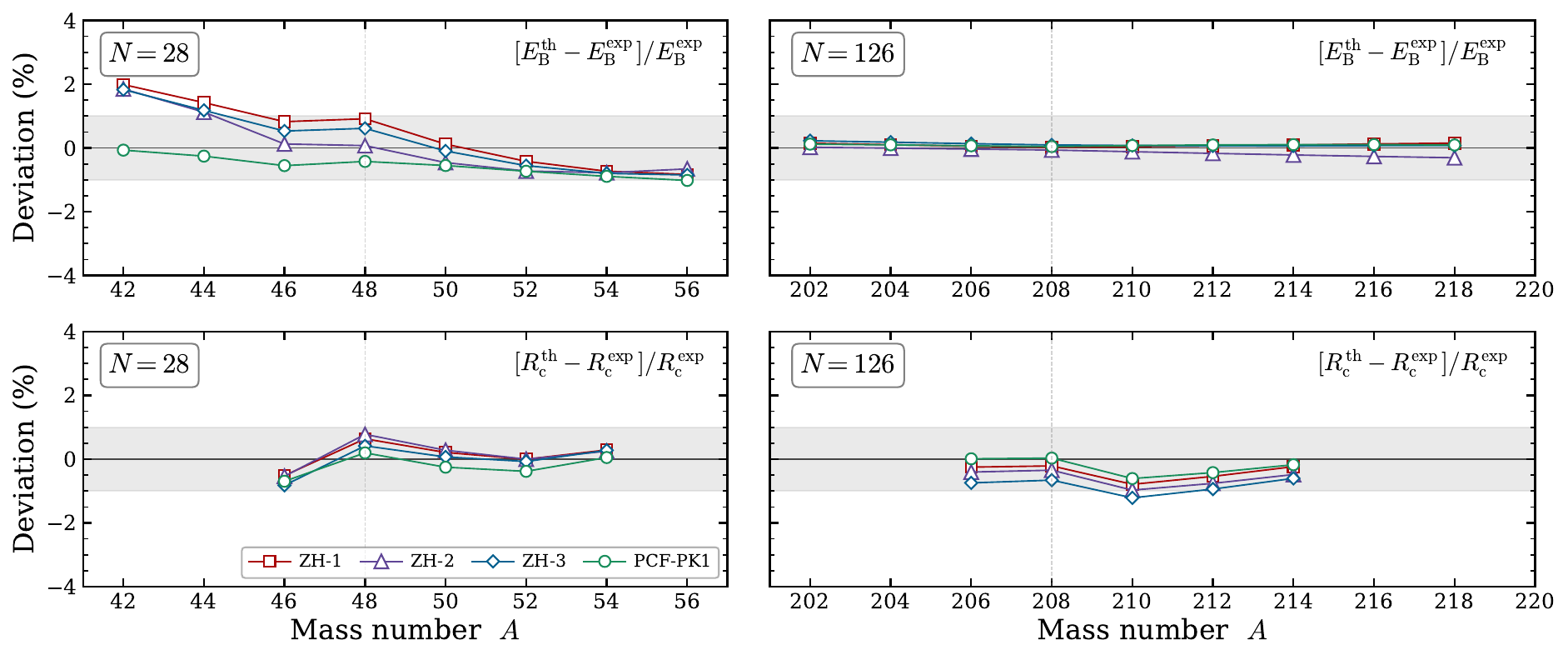}	
	\caption{Same as Fig.~\ref{fig:isotopes}, but for the $N=28$ and $N=126$ isotonic chains.}
	\label{fig:isotones}%
\end{figure*}

\begin{figure*}[tbp]
	\centering 
	\includegraphics[width=0.60\textwidth]{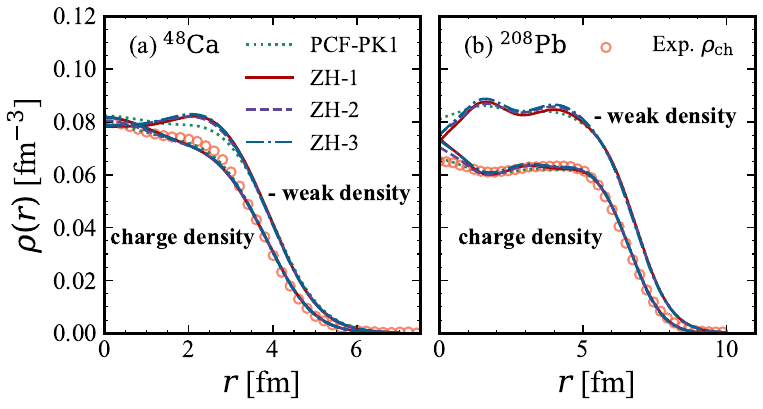}	
	\caption[Charge and weak radial density distributions]{The charge (solid lines) and weak (dashed lines) radial density distributions in $^{48}\mathrm{Ca}$~(a) and $^{208}\mathrm{Pb}$~(b) predicted by ZH-1~(red), ZH-2~(purple), ZH-3~(blue), and PCF-PK1~(green). Experimental charge densities~\cite{DEVRIES1987495} are shown as open circles.} 
	\label{fig:dens_cw}%
\end{figure*}

\begin{figure*}[tbp]
	\centering 
	\includegraphics[width=0.86\textwidth]{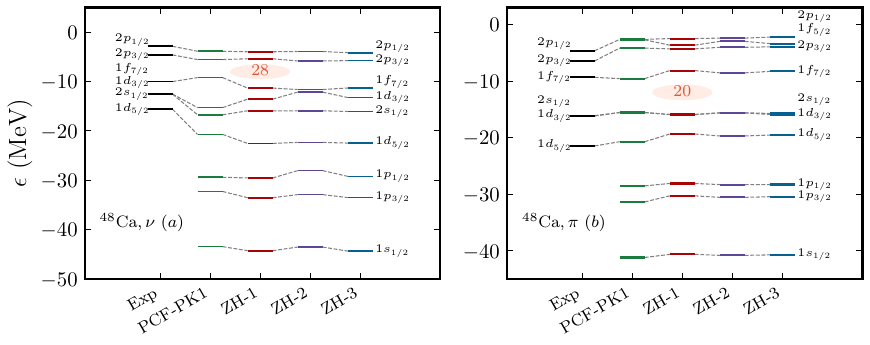}	
	\caption{Neutron (a) and proton (b) single-particle energies of $^{48}$Ca predicted by PCF-PK1, ZH-1, ZH-2, and ZH-3. Experimental data are taken from Ref.~\cite{Schwierz:2007ve}.} 
	\label{fig:SPE_48}%
\end{figure*}

\begin{figure*}[tbp]
	\centering 
	\includegraphics[width=0.9\textwidth]{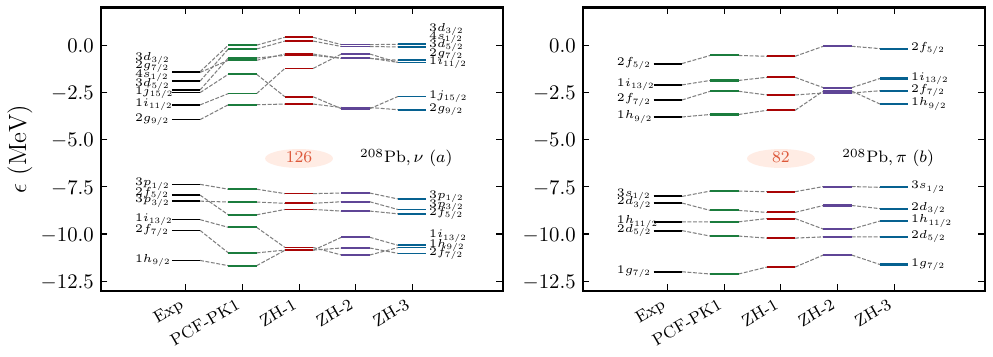}	
	\caption{Same as Fig.~\ref{fig:SPE_48}, but for $^{208}$Pb.} 
	\label{fig:SPE_208}%
\end{figure*} 

\begin{figure*}[tbp]
	\centering 
	\includegraphics[width=0.5\textwidth]{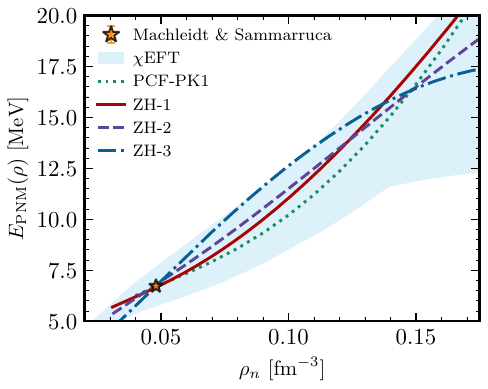}	
	\caption[Pure neutron matter equations of state]{The pure neutron matter equations of state predicted by ZH-1 (red solid), ZH-2 (purple dashed), and ZH-3 (blue dash-dotted) are shown as functions of neutron density $\rho_n$. For comparison, PCF-PK1 is shown as a green dotted line, the combined $\chi$EFT prediction from various many-body calculations~\cite{Huth:2020ozf} is shown as a blue shaded band, and the N$^{3}$LO chiral EFT result of \textit{Machleidt} \& \textit{Sammarruca}~\cite{Machleidt:2024bwl}, $E_{\mathrm{PNM}}(\rho = 0.04805~\mathrm{fm}^{-3})=6.72 \pm 0.08$~MeV, is marked by an orange star.}
	\label{fig:Epnm}%
\end{figure*}

To assess the performance of the ZH EDFs over a broader range of isospin asymmetry, we extend the calculations to the Ca and Pb isotopic chains and to the $N=28$ and $N=126$ isotonic chains. Figures~\ref{fig:isotopes} and~\ref{fig:isotones} show the signed relative deviations of the calculated binding energies and charge radii from the experimental values~\cite{Angeli:2013epw,Wang:2021xhn}. For binding energies, the deviations remain below about $0.7\%$ along the Pb isotopic chain and below about $0.4\%$ along the $N=126$ isotonic chain. For the lighter Ca and $N=28$ chains, most deviations are within about $1\%$--$2\%$, with the largest deviations reaching about $3\%$ in the Ca chain and about $2\%$ in the $N=28$ chain. The larger deviations in the Ca chain are mainly associated with ZH-2, which has the strongest isovector tensor coupling among the three ZH parametrizations. For charge radii, the ZH parametrizations give a reasonable description along the Ca and $N=28$ chains, with most deviations within about $1\%$. Along the Pb isotopic chain, they show a smooth systematic underestimation of $R_c$ by about $0.5\%$--$1.5\%$. A similar but milder trend is found along the $N=126$ isotonic chain, where the deviations are mostly below about $1\%$. These results indicate that the ZH EDFs retain a reasonable overall description of the mass and radius systematics along the extended isotopic and isotonic chains. Note that the isoscalar-scalar and isoscalar-vector channels, together with the pairing interaction, are kept unchanged from PCF-PK1 in the present construction. Further improvement of the global mass and radius description would require a broader calibration of the EDF parameter space.

Some recent attempts to reconcile the PREX-II and CREX results have found that strongly modified isovector sectors may lead to unphysical density oscillations in the nuclear interior~\cite{Reed:2023cap,Salinas:2023qic}.
We therefore examine the radial charge and weak densities predicted by the ZH family. Fig.~\ref{fig:dens_cw} shows the results for $^{48}\mathrm{Ca}$ and $^{208}\mathrm{Pb}$, together with those from PCF-PK1 and the experimental charge-density data~\cite{DEVRIES1987495}. All three ZH EDFs reproduce the charge densities well and exhibit smooth interior behavior, with no spurious oscillations. For $^{48}\mathrm{Ca}$, their predicted weak densities differ from those of PCF-PK1 mainly in the surface region, whereas in $^{208}\mathrm{Pb}$ the differences are very small.

In addition to the density profiles, the robustness of shell structure must also be examined. Ref.~\cite{Kunjipurayil:2025xss} suggested that an excessively strong IVSO potential may induce a level crossing between the neutron $l=6$ SO partners in $^{208}$Pb, potentially affecting the $N=126$ shell closure. Figs.~\ref{fig:SPE_48} and~\ref{fig:SPE_208} display the single-particle spectra predicted by the ZH family and PCF-PK1. Despite the enhanced IVSO strength, the ordering of SO partners near the Fermi surface remains correct and the shell closures are preserved.

Finally, we examine the description of nuclear matter. Fig.~\ref{fig:Epnm} shows the pure neutron matter equation of state $E_{\mathrm{PNM}}$ as a function of neutron density $\rho_n$. Also shown in Fig.~\ref{fig:Epnm} are predictions from chiral effective field theory ($\chi$EFT) calculations~\cite{Huth:2020ozf, Machleidt:2024bwl}. As demonstrated, all three new predictions are in good agreement with the microscopic predictions~\cite{Huth:2020ozf, Machleidt:2024bwl}.

\section{Nonrelativistic reduction and the spin-orbit couplings\label{App:nonrel}}

Neglecting the kinetic, Coulomb and pairing contributions, the energy density of the nuclear system is given by
\begin{equation}\label{eq:enerden}
\begin{aligned}
\mathcal{E} =&
\frac{1}{2}\alpha_{\mathrm{S}}\rho_{\mathrm{S}}^2
+\frac{1}{2}\alpha_{\mathrm{V}}\rho_{\mathrm{V}}^2
+\frac{1}{2}\alpha_{\tau \mathrm{S}}\rho_{\tau \mathrm{S}}^2 
+\frac{1}{2}\alpha_{\tau \mathrm{V}}\rho_{\tau \mathrm{V}}^2
\\
&
-\alpha_{\tau \mathrm{T}}{\mathbf{j}_{\tau \mathrm{T}}^0}^2 
-\alpha_{ \mathrm{T}}{\mathbf{j}_{ \mathrm{T}}^0}^2
-\frac{1}{2} \delta_{\mathrm{S}}\left(\nabla \rho_{\mathrm{S}}\right)^2,
\end{aligned}
\end{equation}
where
\begin{equation}
\begin{aligned}
\rho_{\mathrm{S}}      &= \rho_{\mathrm{S},n} + \rho_{\mathrm{S},p}, 
&\qquad
\rho_{\tau\mathrm{S}}  &= \rho_{\mathrm{S},n} - \rho_{\mathrm{S},p}, \\
\rho_{\mathrm{V}}      &= \rho_{\mathrm{V},n} + \rho_{\mathrm{V},p},
&\qquad
\rho_{\tau\mathrm{V}}  &= \rho_{\mathrm{V},n} - \rho_{\mathrm{V},p}, \\
\mathbf{j}_{\mathrm{T}}^{0}
                       &= \mathbf{j}_{\mathrm{T},n}^{0} + \mathbf{j}_{\mathrm{T},p}^{0},
&\qquad
\mathbf{j}_{\tau\mathrm{T}}^{0}
                       &= \mathbf{j}_{\mathrm{T},n}^{0} - \mathbf{j}_{\mathrm{T},p}^{0}.
\end{aligned}
\end{equation}
The nucleon scalar, vector and tensor densities are defined as
\begin{equation}\label{eq:dens}
\begin{aligned}
\rho_{\mathrm{S},q} & =\sum_\alpha \bar{\psi}_\alpha \psi_\alpha=\sum_\alpha\left[\varphi_\alpha^{ \dagger} \varphi_\alpha-\chi_\alpha^{ \dagger} \chi_\alpha\right], \\
\rho_{\mathrm{V},q} & =\sum_\alpha \bar{\psi}_\alpha \gamma^0 \psi_\alpha=\sum_\alpha\left[\varphi_\alpha^{ \dagger} \varphi_\alpha+\chi_\alpha^{ \dagger} \chi_\alpha\right], \\ 
\mathbf{j}_{\mathrm{T},q}^0 & =\sum_\alpha \bar{\psi}_\alpha i \gamma^0 \boldsymbol{\gamma} \psi_\alpha=\sum_\alpha\left[\varphi^{\dagger}_\alpha i \boldsymbol{\sigma} \chi_\alpha-\chi_\alpha^{\dagger} i \boldsymbol{\sigma}\varphi_\alpha\right],
\end{aligned}
\end{equation}
where $q=p,n$, and $\psi_{\alpha}=(\varphi_{\alpha},\chi_{\alpha})^{\mathrm{T}}$ is the Dirac four-spinor of the occupied single-particle state $\alpha$ for nucleon species $q$, with $\varphi_{\alpha}$ and $\chi_{\alpha}$ denoting its upper and lower components, respectively.

To reveal the implicit SO interaction, we perform a nonrelativistic reduction and re-express $\mathcal{E}$ in terms of standard nonrelativistic densities and currents built from the normalized \emph{classical} wave function $\phi^{\mathrm{cl}}$~\cite{Reinhard:1989nr, Sulaksono:2003re, Sulaksono:2011zz}:
\begin{equation}
\begin{aligned}
\rho_q  &=\sum_\alpha\left|\phi_\alpha^{\mathrm{cl}}\right|^2,\quad\quad\tau_q =\sum_\alpha\left|\nabla \phi_\alpha^{\mathrm{cl}}\right|^2, \\
\boldsymbol{J}_q & =-\frac{i}{2} \sum_\alpha\left[\phi_\alpha^{\mathrm{cl} \mathrm{\dagger}}(\nabla \times \boldsymbol{\sigma}) \phi^{\mathrm{cl}}_{\alpha}-\left(\nabla \times \boldsymbol{\sigma} \phi_\alpha^{\mathrm{cl}}\right)^{\dagger} \phi_\alpha^{\mathrm{cl}}\right],
\end{aligned}
\end{equation}
where $\rho_q$, $\tau_q$, and $\boldsymbol{J}_q$ are nucleon number density, kinetic energy density and SO density, respectively.

 To proceed, we eliminate the lower component of the Dirac four-spinor. For simplification of notation, we omit the state index $\alpha$ in what follows.
Variation of the Lagrangian density yields the single-nucleon Dirac equation
\begin{equation}
\left[\boldsymbol{\alpha} \cdot \boldsymbol{p} + \beta(m + S_q) + V_q^0 - i \beta \boldsymbol{\alpha} \cdot \boldsymbol{T}_q^0\right] \psi = \epsilon_q \psi,
\end{equation}
where the scalar, vector and tensor potentials are defined as
\begin{equation}\label{eq:poten}
\begin{aligned}
&S_q  = \alpha_{\mathrm{S}} \rho_{\mathrm{S}} + \tau_3 \alpha_{\tau \mathrm{S}} \rho_{\tau \mathrm{S}} + \delta_{\mathrm{S}} \Delta \rho_{\mathrm{S}}, \\
&V_q^0  = \alpha_{\mathrm{V}} \rho_{\mathrm{V}} +\tau_3  \alpha_{\tau \mathrm{V}} \rho_{\tau \mathrm{V}} + e \frac{1 - \tau_3}{2} A^0 + \Sigma^r, \\
&\boldsymbol{T}_q^0  = 2\alpha_{\mathrm{T}} \mathbf{j}_{\mathrm{T}}^0 + 2 \tau_3\alpha_{\tau \mathrm{T}} \mathbf{j}_{\tau \mathrm{T}}^0,
\end{aligned}
\end{equation}
with $\tau_3 =1$ and $-1$ for neutrons and protons, respectively. The rearrangement term $\Sigma^r$ is given by 
\begin{equation}
\begin{aligned}
\Sigma^r  = &
\frac{1}{2} \left(
\frac{\partial {\alpha}_{\mathrm{V}}}{\partial \rho_{\mathrm{V}}} \rho_{\mathrm{V}}^{2}
+\frac{\partial {\alpha}_{\mathrm{S}}}{\partial \rho_{\mathrm{V}}} \rho_{\mathrm{S}}^{2}
+\frac{\partial {\alpha}_{\tau \mathrm{S}}}{\partial \rho_{\mathrm{V}}} \rho_{\tau \mathrm{S}}^{2}
+\frac{\partial {\alpha}_{\tau \mathrm{V}}}{\partial \rho_{\mathrm{V}}} \rho_{\tau \mathrm{V}}^{2}
\right)
\\
&
-\frac{\partial {\alpha}_{\mathrm{T}}}{\partial \rho_{\mathrm{V}}} \mathbf{j}_{\mathrm{T}}^{2}
-\frac{\partial {\alpha}_{\tau \mathrm{T}}}{\partial \rho_{\mathrm{V}}} \mathbf{j}_{\tau\mathrm{T}}^{2} .
\end{aligned}
\end{equation}
In Dirac-Pauli representation, the single-nucleon Dirac equation takes the matrix form 
\begin{equation}\label{eq:Diraceq}
\left(\begin{array}{cc}
m+S+V^0 & \boldsymbol{\sigma} \cdot( \boldsymbol{p} -i \boldsymbol{T}^0)  \\
\boldsymbol{\sigma} \cdot (\boldsymbol{p}  +i\boldsymbol{T}^0 )    & -m-S+V^0
\end{array}\right)\binom{\varphi}{\chi} = \epsilon \binom{\varphi}{\chi}.
\end{equation}
From the second row of Eq.~(\ref{eq:Diraceq}), one obtains
\begin{equation}\label{eq:comp}
\chi = \frac{\boldsymbol{\sigma} \cdot (\boldsymbol{p} +i\boldsymbol{T}^0) }{\epsilon+m+S_q-V_q^0}\varphi \equiv \mathcal{B}\boldsymbol{\sigma} \cdot \overrightarrow{\Pi}\varphi.
\end{equation}
with $\overrightarrow{\Pi }= \boldsymbol{p}  +i \boldsymbol{T}^0$ and $\mathcal{B}_q = \left[\epsilon+m+S_q-V_q^0\right]^{-1}$. 

Next, we identify the upper component $\varphi$ with the \emph{classical} wave function $\phi^{\mathrm{cl}}$.
Note that the vector density $\rho_{\mathrm{V}}$ must satisfy the normalization condition, which implies
\begin{equation}
\int d^3 r \bar{\psi} \gamma^0 \psi=\int \mathrm{d}^3 r \varphi^{\dagger}\hat{I} \varphi =1 =\int \mathrm{d}^3 r \phi^{\mathrm{cl }\dagger} \phi^{\mathrm{cl}} ,
\end{equation}
where the norm kernel is
\begin{equation}
 \hat{I} =1+\boldsymbol{\sigma} \cdot \overleftarrow{\Pi} \mathcal{B}^2 \boldsymbol{\sigma} \cdot \overrightarrow{\Pi}.\label{eq:norm}
 \end{equation}
This establishes the transformation between $\varphi$ and $\phi^{\mathrm{cl}}$ as
\begin{equation}\label{eq:trans}
\begin{aligned}
\phi^{\mathrm{cl}} = \hat{I}^{1/2} \varphi, \quad
\varphi = \hat{I}^{-1/2}\phi^{\mathrm{cl}}.
\end{aligned}
\end{equation}

Using these relations, the nonrelativistic limit is obtained by a systematic expansion in the small quantity
\begin{equation}
\mathcal{B}_0 = 1/(2m^{*}),
\end{equation} where
$
m^{*} = m + \alpha_{\mathrm{S}}\,\rho_{\mathrm{S}}
$
denotes the isoscalar Dirac effective mass. In this power counting, $S_q+V_q^{0}$, $\epsilon_q-m$, 
$\tau_{3}\alpha_{\tau\mathrm{S}}\rho_{\tau\mathrm{S}} $ and  $\delta_{\mathrm{S}}\Delta\rho_{\mathrm{S}}$
are treated as leading-order quantities, $\mathcal{O}(1)$, 
and thus $\mathcal{B}$ defined in Eq.~(\ref{eq:comp}) can be 
expanded as 
\begin{equation}
\begin{aligned}
\mathcal{B} =&\,
\Bigl[
2m^* -(S_q+V_q^0)+(\epsilon_q-m)
+2(\tau_3\alpha_{\tau \mathrm{S}}\rho_{\tau \mathrm{S}}
+\delta_{\mathrm{S}}\Delta \rho_{\mathrm{S}})
\Bigr]^{-1}\\
\simeq &\,\mathcal{B}_{0}
      + \mathcal{O}(\mathcal{B}_{0}^{2}).
\end{aligned}
\end{equation}
Taylor expanding the norm operator $\hat{I}^{-1/2}$ then gives
\begin{equation}\label{eq:tayl}
\hat{I}^{-1/2}
\simeq 1
-\frac{1}{2}\left(
\boldsymbol{\sigma}\cdot\overleftarrow{\Pi}\,
\mathcal{B}_{0}^{2}\,
\boldsymbol{\sigma}\cdot\overrightarrow{\Pi}
\right)
+ \mathcal{O}(\mathcal{B}_0^3),
\end{equation}
where $\overrightarrow{\Pi} = \boldsymbol{p}+i\boldsymbol{T}^{0}$ and
$\overleftarrow{\Pi}$ acts to the left.

Substituting Eqs.~(\ref{eq:comp}), (\ref{eq:trans}), and (\ref{eq:tayl}) into the density definitions in Eq.~(\ref{eq:dens}), and keeping terms consistently up to order $\mathcal{B}_{0}^{2}$, one obtains the nonrelativistic limits
\begin{equation}\label{eq:nrdens}
\begin{aligned}
\rho_{\mathrm{V},q} &= \rho_q,\\
\rho_{\mathrm{S},q}
&= \sum_{\alpha}
   \phi_{\alpha}^{\mathrm{cl}\,\dagger}
   \hat{I}^{-1/2}
   \left(
   1
   - \boldsymbol{\sigma}\cdot\overleftarrow{\Pi}\,
     \mathcal{B}_{0}^{2}\,
     \boldsymbol{\sigma}\cdot\overrightarrow{\Pi}
   \right)
   \hat{I}^{-1/2}
   \phi_{\alpha}^{\mathrm{cl}}
\\
&= \rho_q
 - 2\mathcal{B}_{0}^{2}
   \left [
   \tau_q
   - {\nabla}\!\cdot\boldsymbol{J}_q
   - {\nabla}\rho_q\cdot\boldsymbol{T}_q^{0}
   + 2\,\boldsymbol{T}_q^{0}\!\cdot\boldsymbol{J}_q
   + \rho_q(\boldsymbol{T}_q^{0})^{2}
   \right ],\\
\mathbf{j}_{\mathrm{T},q}^{0}
&= \sum_{\alpha}
   \Big[
   \phi_{\alpha}^{\mathrm{cl}\,\dagger}
   \hat{I}^{-1/2}
   \big(
   i\,\boldsymbol{\sigma}\,\mathcal{B}_{0}\,
   \boldsymbol{\sigma}\cdot\overrightarrow{\Pi}
   \big)
   \hat{I}^{-1/2}
   \phi_{\alpha}^{\mathrm{cl}}
   - \textit{c.c.}
   \Big]
\\[0.5mm]
&= \mathcal{B}_{0}\,{\nabla}\rho_q
 - 2\mathcal{B}_{0}\,\rho_q \,\boldsymbol{T}_q^{0}
 - 2\mathcal{B}_{0}\,\boldsymbol{J}_q.
\end{aligned}
\end{equation}

Replacing the relativistic densities and currents in Eq.~(\ref{eq:enerden}) by their nonrelativistic limits in Eq.~(\ref{eq:nrdens}), and expanding consistently up to order $\mathcal{B}_{0}^{2}$, we obtain the nonrelativistic reduction of the covariant potential energy density:
\begin{equation}\label{eq:NREden}
\begin{aligned}
\mathcal{E} =&\,
\left(\frac{1}{2}\alpha_{\mathrm{S}}
    + \frac{1}{2}\alpha_{\mathrm{V}}\right)\rho^{2}
+ \left(\frac{1}{2}\alpha_{\tau \mathrm{S}}
    + \frac{1}{2}\alpha_{\tau \mathrm{V}}\right)\tilde{\rho}^{2}
- \frac{1}{2}\delta_{\mathrm{S}}({\nabla}\rho)^{2}
\\
&
- 2 \mathcal{B}_0^{2} \alpha_{\mathrm{S}}\, \rho\,\tau
- 2 \mathcal{B}_0^{2} \alpha_{\tau \mathrm{S}}\, \tilde{\rho}\,\tilde{\tau}
 - \alpha_{\mathrm{T}} \mathcal{B}_0^{2}({\nabla}\rho)^{2}
  - \alpha_{\tau \mathrm{T}} \mathcal{B}_0^{2}({\nabla} \tilde{\rho})^{2}
\\[1mm]
&
  - 4 \alpha_{\mathrm{T}} \mathcal{B}_0^{2} \boldsymbol{J}^{2}
  - 4 \alpha_{\tau \mathrm{T}} \mathcal{B}_0^{2} \tilde{\boldsymbol{J}}^{2}
  - 2 \mathcal{B}_0^{2} \alpha_{\tau \mathrm{S}}^{\prime}
    \tilde{\rho}\,{\nabla}\rho \cdot \tilde{\boldsymbol{J}}
\\[1mm]
& + \left(
   - 2 \mathcal{B}_0^{2} \alpha_{\mathrm{S}}^{\prime} \rho
   - 2 \mathcal{B}_0^{2} \alpha_{\mathrm{S}}
   + 4 \mathcal{B}_0^{2} \alpha_{\mathrm{T}}
   \right){\nabla}\rho \cdot \boldsymbol{J}
\\[1mm]
& + \left(
   - 2 \mathcal{B}_0^{2} \alpha_{\tau \mathrm{S}}
   + 4 \mathcal{B}_0^{2} \alpha_{\tau \mathrm{T}}
   \right){\nabla}\tilde{\rho} \cdot \tilde{\boldsymbol{J}},
\end{aligned}
\end{equation}
where $\tilde{\rho} = \rho_{n} - \rho_{p}$, $\tilde{\tau} = \tau_{n} - \tau_{p}$, and $\tilde{\boldsymbol{J}} = \boldsymbol{J}_{n} - \boldsymbol{J}_{p}$. Primes on the couplings denote derivatives with respect to the (isoscalar) density. We note that different choices for expanding $\mathcal{B}$ exist in the
literature~\cite{Reinhard:1989nr,Sulaksono:2003re,Sulaksono:2011zz}, but the results differ only at higher order.

\begin{figure*}[tbp]
	\centering 
	\includegraphics[width=0.7\textwidth]{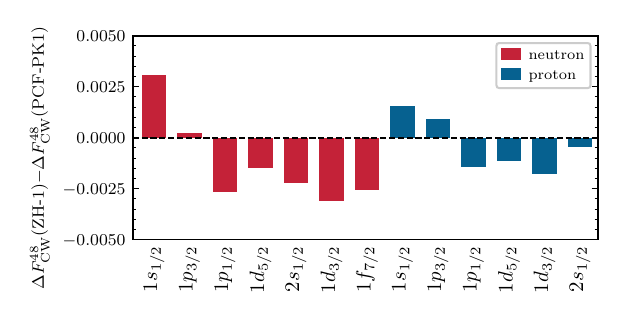}	
	\caption{Relative changes of the single-orbital contributions to the weak-charge form factor difference $\Delta F_{\mathrm{CW}}^{48}$ in $^{48}$Ca, obtained with ZH-1 relative to PCF-PK1. Neutron and proton orbitals are indicated in red and blue, respectively.}
	\label{fig:SO_48}%
\end{figure*} 

By comparing Eq.~(\ref{eq:NREden}) with nonrelativistic Skyrme EDF~\cite{Yue:2024srj}, we identify the SO part of the functional as
\begin{equation}
\mathcal{E}_{\mathrm{SO}} 
= \frac{b_{\mathrm{IS}}}{2}{\nabla}\rho\!\cdot\!\bm{J}
 + \frac{b_{\mathrm{IV}}}{2}{\nabla}\tilde{\rho}\!\cdot\!\bm{\tilde{J}}
 + \frac{b_{\mathrm{SV}}}{2}{\nabla}\rho\!\cdot\!\bm{\tilde{J}},
\end{equation}
where the strength parameters of the isoscalar and isovector
SO interactions are
\begin{equation}
\begin{aligned}
&b_{\mathrm{IS}} = 8 \mathcal{B}_0^2 \alpha_{\mathrm{T}} -4\mathcal{B}_0^2 \alpha_{\mathrm{S}}-4\mathcal{B}_0^2 \alpha_{\mathrm{S}}^{\prime}\rho,\\
&b_{\mathrm{IV}} = 8 \mathcal{B}_0^2 \alpha_{\tau \mathrm{T}} -4\mathcal{B}_0^2 \alpha_{\tau \mathrm{S}}.\\
\end{aligned}
\end{equation}
Due to the density dependence of $\alpha_{\tau \mathrm{S}}$, an additional mixed term $\nabla \rho\cdot\tilde{\boldsymbol{J}}$ appears in the
nonrelativistic reduction, and we define the corresponding strength as
\begin{equation}
b_{\mathrm{SV}} = -4\mathcal{B}_0^2\alpha_{\tau \mathrm{S}}^{\prime}\tilde{\rho}.
\end{equation}

The nonrelativistic limit of the single-particle Hamiltonian entering into the relativistic Hartree equation can be derived either through variations of Eq.~(\ref{eq:NREden}) or directly by a nonrelativistic reduction of the Dirac Hamiltonian as in Ref.~\cite{Reinhard:1989nr}. Up to order $\mathcal{B}_0^2$, both approaches lead to 
\begin{equation}
{h}_q^{\mathrm{eff}}=\boldsymbol{p}\mathcal{B}_0\boldsymbol{p}+U_q+i \boldsymbol{W}_q \cdot(\boldsymbol{\sigma} \times {\nabla}),\quad q=n, p
\end{equation}
where the effective central potential reads
\begin{equation}\label{eq:U_q}
\begin{aligned}
U_q=&
\left(\alpha_{\mathrm{S}}+\alpha_{\mathrm{V}}\right) \rho
+\tau_3\left(\alpha_{\tau \mathrm{S}}+\alpha_{\tau \mathrm{V}}\right) \tilde{\rho}
\\
&
+\frac{1}{2}\left(\alpha_{\mathrm{S}}^{\prime}+\alpha_{\mathrm{V}}^{\prime}\right) \rho^2
+\frac{1}{2}\left(\alpha_{\tau \mathrm{S}}^{\prime}+\alpha_{\tau \mathrm{V}}^{\prime}\right) {\tilde{\rho}}^2
\\
&
-2 \mathcal{B}_0^2 \alpha_{\mathrm{S}} \tau
-2 \tau_3 \mathcal{B}_0^2 \alpha_{\tau \mathrm{S}} \tilde{\tau}
-2 \mathcal{B}_0^2  \alpha_{\mathrm{S}}^{\prime} \rho \tau
-2 \mathcal{B}_0^2 \alpha_{\tau \mathrm{S}}^{\prime} \tilde{\rho} \tilde{\tau}
\\
&
+\left( 2 \mathcal{B}_0^2 \alpha_{\mathrm{T}} +\delta_{\mathrm{S}} \right)\nabla^2 \rho
+2 \tau_3 \mathcal{B}_0^2  \alpha_{\tau \mathrm{T}} \nabla^2 \tilde{\rho}
 \\
& -\frac{b_{\mathrm{IS}}}{2} \nabla \cdot \boldsymbol{J}-\tau_3\frac{b_{\mathrm{IV}}+\tau_3 b_{\mathrm{SV}}}{2}\nabla  \cdot \tilde{\boldsymbol{J}}.
\end{aligned}
\end{equation}
The corresponding SO potential is
\begin{equation}\label{eq:Wq}
\begin{aligned}
\boldsymbol{W}_q
=&
\frac{b_{\mathrm{IS}}+\tau_3 b_{\mathrm{SV}}}{2} \nabla \rho
+\tau_3 \frac{b_{\mathrm{IV}}}{2} \nabla \tilde{\rho}
-8 \mathcal{B}_0^2\alpha_{\mathrm{T}}  \boldsymbol{J}
-8 \tau_3 \mathcal{B}_0^2 \alpha_{\tau \mathrm{T}} \tilde{\boldsymbol{J}}.\\
\end{aligned}
\end{equation}

For the EDFs used in this study, the value of $b_{\mathrm{SV}}$ is
numerically small (see Table~\ref{tab:so} in the main text), so the SO effects are dominated by $b_{\mathrm{IS}}$ and $b_{\mathrm{IV}}$.
As Eqs.~(\ref{eq:U_q}) and (\ref{eq:Wq}) indicate, $b_{\mathrm{IS}}$ and $b_{\mathrm{IV}}$ affect not only the SO potential $\bm{W}_q$ but also the central mean field $U_q$. To clarify which contribution primarily drives the IVSO effect on the charge-weak form factor difference
$\Delta F_{\mathrm{CW}}^{48}$ in $^{48}$Ca, we evaluated the single-orbital contributions to the deviation between the
predictions of ZH-1 and PCF-PK1. As shown in Fig.~\ref{fig:SO_48}, the contributions from individual orbitals 
are relatively uniform, in contrast to Ref.~\cite{Kunjipurayil:2025xss}, 
which reported a dominant effect from the neutron $1f_{7/2}$ orbital via the 
SO potential.  
Our results indicate instead that the main reduction in 
$\Delta F_{\mathrm{CW}}^{48}$ due to the strong isovector tensor coupling originates from changes in the effective 
central potential.

\section{\texorpdfstring{Correlations of $\Delta F_{\mathrm{CW}}$ and $\Delta r_{\mathrm{np}}$ with the symmetry-energy slope at different densities}{Correlations of Delta FCW and Delta rnp with the symmetry-energy slope at different densities}\label{App:correlation}}
 \begin{figure*}[tbp]
	\centering 
	\includegraphics[width=0.66\textwidth]{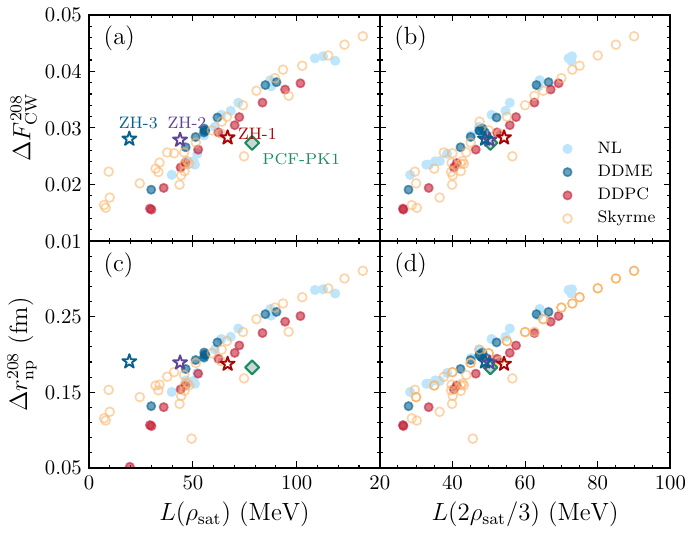}	
	\caption[Correlations with the symmetry-energy slope in $^{208}$Pb]{Data-to-data correlations of the charge-weak form factor difference $\Delta F_{\mathrm{CW}}$ in panels (a,b) and the neutron skin thickness $\Delta r_{\mathrm{np}}$ in panels (c,d) of $^{208}$Pb with the density-slope parameter of the symmetry energy, $L(\rho)$, evaluated at $\rho=\rho_{\mathrm{sat}}$ (left panels) and $\rho=2\rho_{\mathrm{sat}}/3$ (right panels).}
\label{fig:R-L_208}
\end{figure*} 
\begin{figure*}[tbp]
	\centering 
	\includegraphics[width=0.66\textwidth]{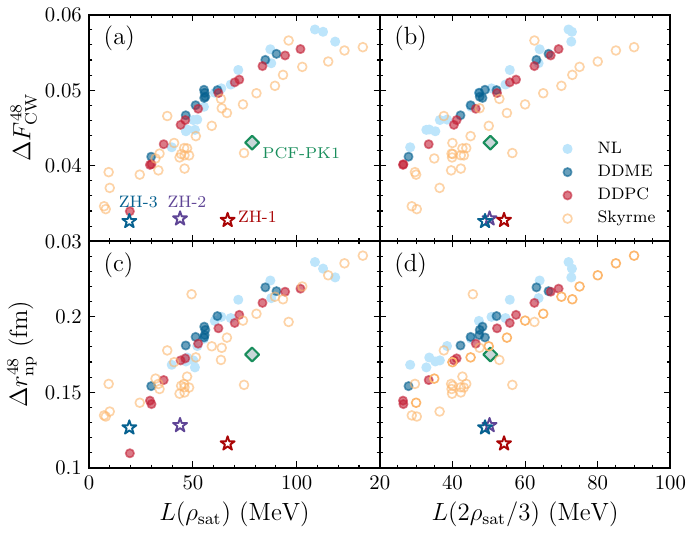}	
	\caption{Same as Fig.~\ref{fig:R-L_208}, but for $^{48}$Ca.}
	\label{fig:R-L_48}%
\end{figure*}

We show in Figs.~\ref{fig:R-L_208} and~\ref{fig:R-L_48} the data-to-data correlations of the charge-weak form factor difference $\Delta F_{\mathrm{CW}}$ (upper panels) and the neutron skin thickness $\Delta r_{\mathrm{np}}$ (lower panels) in $^{208}$Pb and $^{48}$Ca with the symmetry-energy slope parameter $L(\rho)$ evaluated at $\rho_{\mathrm{sat}}$ and $2\rho_{\mathrm{sat}}/3$.
In addition to the ZH family and PCF-PK1, the comparison includes a broad set of relativistic and nonrelativistic EDFs.
The relativistic EDFs comprise 16 nonlinear meson exchange (NL) models (FSU~\cite{Todd-Rutel:2005yzo,Chen:2014sca,Piekarewicz:2010fa} and NL3 families~\cite{Lalazissis:1996rd,Horowitz:2002mb}, IOPB-1~\cite{Miyatsu:2023lki}, IU-FSU~\cite{Fattoyev:2010rx}, $\mathrm{IU}$-$\mathrm{FSU}^{*}$~\cite{Agrawal:2012rx}, FSU-Garnet~\cite{Miyatsu:2023lki}, BigApple~\cite{Fattoyev:2020cws}), 10 density-dependent meson-exchange (DDME) models (DD~\cite{Typel:2005ba}, DD2~\cite{Typel:2009sy}, PK-DD~\cite{Long:2003dn}, TW-99~\cite{Typel:1999yq}, DDME1~\cite{Niksic:2002yp}, DDME2~\cite{Lalazissis:2005de}, DDMEJ family~\cite{Vretenar:2003qm}), and 13 density-dependent point-coupling (DDPC) models (DD-PCX~\cite{Yuksel:2019dnp}, DDPC-PREX, DDPC-CREX, DDPC-REX~\cite{Yuksel:2022umn}, DD-PC1~\cite{Niksic:2008vp}, DD-PCJ family~\cite{Yuksel:2021fht}).
For completeness, 38 Skyrme EDFs are also included, such as SIII, SIV, SV, SVI~\cite{Beiner:1974gc}, SLy230a, SLy230b, SLy4, SLy5, SLy8~\cite{Chabanat:1997qh, Chabanat:1997un}, SAMi~\cite{Roca-Maza:2012dhp}, SGI, SGII~\cite{vanGiai:1981zz}, SkM~\cite{Krivine:1980kzz}, SkM$^{*}$~\cite{Bartel:1982ed}, Ska~\cite{Kohler:1976fgx}, MSk1, MSk2~\cite{Tondeur:2000bd}, MSk7~\cite{Goriely:2001zz}, BSk1~\cite{Samyn:2002bbs}, BSk2~\cite{Goriely:2002xz}, and the SkyLc family~\cite{Yue:2021yfx}.

Although both relativistic and nonrelativistic conventional EDFs exhibit a strong correlation between $\Delta r_{\mathrm{np}}^{208}$ (or $\Delta F_{\mathrm{CW}}^{208}$) and the slope parameter at saturation density, $L(\rho_{\mathrm{sat}})$, the ZH family shows a clear deviation. Despite large variations in $L(\rho_{\mathrm{sat}})$, the three ZH EDFs predict nearly identical values of $\Delta r_{\mathrm{np}}^{208}$ and $\Delta F_{\mathrm{CW}}^{208}$.
This behavior indicates that the inclusion of a strong isovector tensor coupling can disrupt the conventional $\Delta r_{\mathrm{np}}^{208}$-$L(\rho_{\mathrm{sat}})$ correlation.
Nevertheless, the correlation between $\Delta r_{\mathrm{np}}^{208}$ (or $\Delta F_{\mathrm{CW}}^{208}$) and $L(2\rho_{\mathrm{sat}}/3)$ remains robust. For $^{48}$Ca, however, $\Delta r_{\mathrm{np}}^{48}$ (or $\Delta F_{\mathrm{CW}}^{48}$) is also sensitive to the IVSO interaction because of the eight SO unpaired neutrons in the $1f_{7/2}$ orbital. Consequently, as illustrated in Fig.~\ref{fig:R-L_48}, the ZH family exhibits a clear departure from the usual correlation between $\Delta r_{\mathrm{np}}^{48}$ (or $\Delta F_{\mathrm{CW}}^{48}$) and $L(2\rho_{\mathrm{sat}}/3)$.

\FloatBarrier

\section*{Acknowledgements}

ZZ would like to thank Qiang Zhao and Peng-Wei Zhao for valuable discussions on the covariant DDPC density functional, and Bao-Jun Cai and Rui Wang for discussions on the manuscript. This work was supported in part by the National Natural Science Foundation of China under Grant Nos.~12235010, 12575137 and 12147101, the National SKA Program of China under Grant No.~2020SKA0120300, and the Science and Technology Commission of Shanghai Municipality under Grant No.~23JC1402700.

\bibliographystyle{elsarticle-num}
\bibliography{reference}

\end{document}